\documentclass[prb,aps,twocolumn,superscriptaddress,showpacs,floatfix]{revtex4-1}
\usepackage[pdftex]{graphicx}
\usepackage{amsmath}
\usepackage{amssymb}

\newcommand{\vv}[1]{\mathbf{#1}}
\newcommand{\up}{\uparrow}
\newcommand{\down}{\downarrow}
\newcommand{\la}{\langle}
\newcommand{\ra}{\rangle}
\DeclareMathOperator{\Tr}{Tr}

\begin{document}

\title{N\'eel temperature and thermodynamics of the half-filled 3D Hubbard model by Diagrammatic Determinant Monte Carlo}

\author{E. Kozik}
\affiliation{Centre de Physique Th\'eorique, Ecole Polytechnique, 91128 Palaiseau Cedex, France}

\author{E. Burovski}
\affiliation{Physics Department, Lancaster University, Lancaster LA1 4YB, UK}

\author{V.W. Scarola}
\affiliation{Department of Physics, Virginia Tech, Blacksburg, Virginia 24061, USA}

\author{M. Troyer}
\affiliation{Theoretische Physik, ETH Zurich, 8093 Zurich, Switzerland}

\begin{abstract}
We study thermodynamics of the 3D Hubbard model at half filling on approach to the N\'eel transition by means of large-scale unbiased Diagrammatic Determinant Monte Carlo simulations. We obtain the transition temperature in the strongly correlated regime, as well as temperature dependence of energy, entropy, double occupancy, and the nearest-neighbor spin correlation function. Our results improve the accuracy of previous unbiased studies and present accurate benchmarks in the ongoing effort to realize the antiferromagnetic state of matter with ultracold atoms in optical lattices. 
\end{abstract}

\maketitle

\section{Introduction} 


The Hubbard model\cite{Hubbard} of interacting fermions in a solid is a centerpiece of modern condensed matter physics. 
It is conventionally defined by restricting the motion of electrons in a crystalline solid to a single band, and simplifying the screened long-range Coulomb interactions between electrons to an on-site repulsion: 

\begin{equation}
\widehat{H} = -t\sum_{\langle \vv{x}\vv{y} \rangle \sigma} c^\dagger_{\vv{x}\sigma} c_{\vv{y}\sigma} + \mathrm{h.c.} %
+ U\sum_{\vv{x}} n_{\vv{x}\uparrow}   n_{\vv{x}\downarrow} -\mu\sum_{\vv{x}\sigma} n_{\vv{x}\sigma} \;,
\label{Hubbard}
\end{equation}
where $\sigma=\uparrow, \downarrow$, $c^\dagger_{\vv{x}\sigma}$ creates a fermion on a site $\vv{x}$, $n_{\vv{x}\sigma} = c^\dagger_{\vv{x}\sigma} c_{\vv{x}\sigma}$, the summation in the first term runs over the nearest-neighbor sites of the simple cubic lattice,  $t$ is the hopping amplitude and $U>0$ is the onsite repulsion. Though remarkably simple in appearance, the model has been used to study a wealth of intriguing quantum many-body phenomena that are due to electron correlations in solids, such as interaction-driven insulators \cite{MI_review}, quantum magnetism and high-temperature superconductivity \cite{Anderson97}. However, despite more than a half of century of intensive investigation, the physics of the model is still not completely understood.  

The most challenging yet the most interesting regime is the intermediate regime with interaction comparable to half the bandwidth $U \sim z t$, where $z$ is the number of nearest neighbors of a site.  This regime offers no small parameter to start a controllable analytic theory.  Furthermore, exact analytic solutions are only accessible either in one spatial dimension  \cite{1D_Hubbard} or an infinite\cite{infD_Hubbard} number of spatial dimensions. Substantial progress has been possible with the development of efficient quantum Monte Carlo (QMC) methods (for a recent review, see Ref.~\onlinecite{MC_review}) accompanied by advances in computer technology. Although for generic \textit{bosonic} systems virtually any equilibrium property can nowadays be calculated by QMC with a controlled high accuracy \cite{bosonic_emulator}, systematic-error-free simulations of correlated \textit{fermions} have been limited to a handful of special cases due to the negative sign problem \cite{sign}. The sign problem manifests itself as an exponential scaling of the simulation time with the system volume and inverse temperature making it practically impossible to obtain any reliable information about the system in the thermodynamic (TD) limit. Although in some cases the sign problem can be completely eliminated by choosing a system-specific representation, its general solution is almost certainly not possible \cite{sign}.  

A major step toward understanding strongly correlated systems has been the experimental realization of the Hubbard model with ultracold atomic gases loaded into optical lattices (for recent reviews see Refs. \onlinecite{lattice_review1, lattice_review2}). These systems offer substantial control over the Hamiltonian.  As a result, these experiments can serve as emulators of quantum many-body systems, which allow the accurate study of a given model in a range of parameters inaccessible by analytic and numeric techniques\cite{hofstettar2002}. The recent experimental observation of Mott physics \cite{Esslinger08, Bloch08} in the Hubbard model is a major milestone along these lines. The next crucial step would be a realization of the antiferromagnetic (AFM) transition and the N\'eel state in the Hubbard model, which requires a substantial effort in reaching lower temperatures, controlling the equilibration rates, developing new probing techniques, etc. In addition to these inherent challenges, there is also a fundamental problem related to thermometry in ultracold-atom systems since they are insulated from the environment.  Isolated systems such as these are thus, by their nature, better characterized by entropy, rather than temperature. Moreover, probes have to be calibrated in the relevant regime and the results obtained with the setup validated against available benchmarks. For these purposes, reliable and accurate numeric results for fermionic systems are indispensable because they ultimately allow a full quantitative understanding, as was recently demonstrated by the example of a bosonic optical-lattice emulator \cite{bosonic_emulator}.

This work provides reliable benchmarks for the realization of the N\'eel state in optical lattices as well as for new theoretical methods.  We focus on the special case of the half-filled  ($\langle n_{\vv{x}\sigma} \rangle = 1/2$, or equivalently $\mu=U/2$) three dimensional (3D) Hubbard model \eqref{Hubbard} on a simple cubic lattice. The case of half filling is special due to the $SU(2)$ symmetry of the Hamiltonian, which is ultimately broken by the N\'eel state, making magnetism the leading instability. In the limit $U/t\ll 1$, the effects of the interaction can be studied perturbatively. In the opposite limit, $U/t\gg 1$, Eq.\ \eqref{Hubbard} reduces to the AFM Heisenberg model with $J\sim t^2/U$. While there is no doubt that in the strongly correlated regime the ground state of the half-filled model \eqref{Hubbard} is AFM, mapping out the finite-temperature phase diagram and studying thermodynamics of the system is extremely challenging \cite{Staudt2000, Kent2006, Joerdens, Fuchs, Loh}. To this end, we employ the unbiased continuous-time determinantal diagrammatic Monte Carlo (DDMC) technique \cite{DDMC_njp2006}, which produces numerically exact (up to a known statistical error bar) results for a finite-size system, and which is free from the fermionic sign problem at half filling on bipartite lattices, allowing a reliable extrapolation of results to the thermodynamic limit. 

We use DDMC to determine several critical properties of the Hubbard model with high control and accuracy and compare with high-temperature series expansions\cite{Oitmaa} (HTSE) where possible. We study the range of on-site interaction $4 \leq U/t \leq 8$, where the critical temperature $T_N$ of the AFM transition is expected to reach its maximum \cite{Staudt2000} . Results in this regime are vital to optical lattice emulator efforts offering experiments their best chance of observing the AFM phase. We obtain the critical temperature $T_N$ and compute important thermodynamic properties of the model---the energy and the entropy, as well as two optical lattice observables: double occupancy and the nearest-neighbor spin-spin correlation function---in the paramagnetic phase as a function of temperature down to $T_N$. A number of previous works using different unbiased approaches studied $T_N$ (by the determinant quantum Monte Carlo (DQMC)\cite{Staudt2000} and the Dynamical Cluster Approximation (DCA)\cite{Kent2006}) and the thermodynamic properties in question (by DDMC\cite{Joerdens}, a combination of DCA and DDMC\cite{Fuchs}, and by DQMC\cite{Loh}) in this regime. Our work improves the accuracy of the previous results at half filling and provides the most accurate to date values of the critical temperature $T_N$ and the entropy at the critical point $S_N$ (summarized in the Table.~\ref{table:TNs}).  Accurate knowledge of $T_N$ is particularly important for determining the critical entropy $S_N$ since $S(T)$ is a steep function near the transition, so that the error bar of $S_N$ is mainly due to the uncertainty in $T_N$. The values of $S_N$ are required for an experimental realization for the AFM state. As was noted in Ref.~\onlinecite{Fuchs}, close to the transition, the temperature dependence of the nearest-neighbor spin-spin correlation function $\langle S^{z}_\mathbf{x} S^{z}_{\mathbf{x} + \mathbf{e}_i} \rangle$ ($\mathbf{e}_i$ is the unit vector in the direction $i$) is significantly more pronounced than that of the double occupancy $\langle n_{\mathbf{x}\uparrow} n_{\mathbf{x}\downarrow} \rangle$ of a lattice site making measurements of the nearest-neighbor spin-spin correlations \cite{Bloch2010, Esslinger2011} more suitable for accurate thermometry in this regime. Our results for the spin-spin correlation function can be used for calibration of such a thermometer.  

For over a decade, the DQMC study of the phase diagram of the half-filled 3D Hubbard model by Staudt \textit{et al.} \cite{Staudt2000} has been the main reference for $T_N$ in the correlated regime. Representing the state of the art at that time, Ref.~\onlinecite{Staudt2000} provides a comprehensive comparison of the DQMC data for the N\'eel temperature with those of preceding QMC simulations and approximate theories, e.g., DMFT \cite{DMFT}. We shall not reproduce this comparison here and refer the reader to a review of results for $T_N$ predating Ref.~\onlinecite{Staudt2000}. 

The simulation method of Staudt \textit{et al.} is based on a discrete Hubbard-Stratonovich decoupling in the Hubbard interaction term which requires discretizing the imaginary-time interval $0<\tau<\beta=1/T$ into a finite number of steps of size $\Delta \tau$ thereby introducing a systematic error. Hence, in addition to the standard extrapolation of the finite-size DQMC data to the TD limit, one has to perform an extrapolation with respect to $\Delta \tau \to 0$. Such a double extrapolation is rather laborious. In practice $\Delta \tau$ is usually fixed at a value which is large enough to allow efficient simulations, yet, according to Ref.~\onlinecite{Staudt2000}, such that ``the results are not significantly affected by the extrapolation $\Delta \tau \to 0$''. In the absence of an explicit extrapolation, the degree of control over systematic errors can be questioned. This is where our approach is a major improvement over that of Staudt \textit{et al.} The DDMC technique is formulated directly in continuous imaginary time.  Therefore finite-size corrections are the only systematic error we have to eliminate in our approach. The cost of the absence of the additional systematic error is the computation complexity of the DDMC, which scales as $[\beta U]^3 \Omega^3$ versus the linear in $[\beta/ \Delta \tau] \Omega^3$ scaling of the DQMC of Ref.~\onlinecite{Staudt2000}, for lattices with $\Omega=L^3$ sites. As a result, we are limited by the values of interaction $U \leq 8$, whereas Staudt \textit{et al.} were able to study the model up to $U=12$.

We have also been able to identify the transition itself with a considerable improvement in accuracy over Ref.~\onlinecite{Staudt2000}.  Long-range AFM order in the system causes divergence of the magnetic structure factor 
\begin{equation}
S(\vv{Q}) = \frac{1}{L^3} \sum_{\vv{x}\vv{y}} e^{i\vv{Q}(\vv{x}-\vv{y})} \left\langle s_z(\vv{x}) s_z(\vv{y}) \right\rangle \;,
\label{AFM}
\end{equation}
where $s_z(\vv{x}) = ( n_{\vv{x}\uparrow} -n_{\vv{x}\downarrow} )/2$, and $\vv{Q}=\left(\pi,\pi,\pi \right)$ is the AFM wave vector, so that $S(\vv{Q})/L^3$ is related to the magnetization $m$ in the TD limit, $\lim_{L\to \infty}S(\vv{Q})/L^3=m^2$. In Ref.~\onlinecite{Staudt2000}, the transition temperature is found as the point at which $m$ starts to noticeably deviate from zero, while the magnetization is obtained from a finite-size extrapolation of $ S(\vv{Q}) $ with respect to $L \to \infty$.  Here the scaling power of the finite-size corrections was used as a fitting parameter. An additional (indirect) probe of the transition used by Staudt \textit{et al.} is the peak in the dependence of the specific heat on temperature. In contrast, we use a much more accurate method to determine the critical point.  We find the critical point using a finite-size scaling analysis of $ S(\vv{Q})$ in combination with the technique of Binder crossings \cite{Binder1981}. This approach allows us to get a reliable and accurate value of $ T_N $ by making use of the known scaling law of the magnetic structure factor at the critical point, $ S(\vv{Q}) \propto 1/L^{-2+\eta} $, where $ \eta $ is the anomalous dimension of the particular universality class, so that the quantity $ S(\vv{Q}) L^{-2+\eta} $ becomes scale-invariant at the critical temperature up to non-universal corrections, which we also take into account. Since the N\'eel transition is breaking the $SU(2)$ symmetry of the Hubbard Hamiltonian at half-filling, the universality class is that of the 3D Heisenberg model, which provides the critical exponents for the finite-size scaling analysis. 

In the context of previous calculations of the N\'eel temperature, it is worth noting the work by Kent \textit{et al.} \cite{Kent2006}, where $T_N$ was found from calculations using DCA. Although this work did not lead to any improvement of the precision claimed by Staudt \textit{et al.}, it was shown that the DCA can be used to determine the critical point with a controlled accuracy based on significantly smaller clusters of only $\sim 50$ lattice sites versus up to $\sim 1000$ sites in Ref.~\onlinecite{Staudt2000} and in this work. In addition, good agreement with the values of $T_N$ of Ref.~\onlinecite{Staudt2000} suggested that potential systematic error of the $\Delta \tau \to 0$ extrapolation in Ref.~\onlinecite{Staudt2000} is likely to be small. In the range of interaction $U/t \geq 6$, our results for $T_N$ agree within the error bars with those of Refs.~\onlinecite{Staudt2000, Kent2006} implicitly confirming this. At smaller $U$, however, we find somewhat lower values of $T_N$. Moreover, at $U/t=4$, being unable to reach significantly low temperatures to accurately infer the critical point, we can only claim an upper bound from our finite-size-scaling analysis, $T_N/t < 0.17$, which is already lower than the values claimed in Refs.~\onlinecite{Staudt2000, Kent2006}. The reason for the discrepancy is likely to be the long-range character of correlations at smaller coupling, which can be missed in simple finite-size extrapolation schemes based on data for insufficiently large systems. 

The thermodynamics of the Hubbard model near the N\'eel transition in connection with its experimental realization has been the focus of a number of recent studies \cite{deleo2008,scarola2009,Joerdens, Gorelik2010, DeLeo,Fuchs, Loh, DGammaA, Gorelik2012}. The DMFT results \cite{Gorelik2010, Gorelik2012} emphasize the role of double occupancy in detecting the build up of AFM correlations. However, its dependence on temperature near the N\'eel transition is relatively flat in the regime where $T_N$ is maximal, as observed in unbiased (extrapolated to the TD limit) DCA calculations \cite{Fuchs}. Fuchs \textit{et al.} \cite{Fuchs} obtained the energy and the entropy down to $T_N$ in this regime as well as the equation of state away from half filling, which allowed to get an estimate of the entropy at the transition in a realistic harmonically trapped system. A study\cite{Loh} of the system using the same DQMC method as Ref.~\onlinecite{Staudt2000} arrived at an agreement with the results and conclusions of Fuchs \textit{et al.}. 

The DDMC simulation method used in this work is not capable of capturing thermodynamics of the Hubbard model away from half filling due to the pronounced negative sign problem. However, exactly at half filling, it has certain advantages over DCA and DQMC. As discussed above, in order to claim unbiased results in the TD limit within DQMC one has to resort to a double extrapolation, $\Delta \tau \to 0, L \to \infty$.  DDMC is formulated directly in continuous time allowing a more reliable extrapolation to the TD limit. Modern efficient solvers for DCA also work in continuous time, but the cluster sizes amenable to simulation in the regime of interest are typically less then $100$. However, in DCA the clusters are embedded in a self-consistently defined medium, which largely improves the convergence to the TD limit. In practice the finite-size dependence for the accessible clusters is notably larger than that in DDMC \cite{Fuchs}. Hence, at half filling and $U/t \sim 8$, where $T_N$ and $S_N$ are expected to reach their maxima, DDMC currently allows to obtain the most reliable benchmarks for the thermodynamic quantities of interest.

The paper is organized as follows. In Sec.~\ref{sec:method}, we discuss the simulation method outlining the general formulation of the DDMC technique and its application to calculating the specific observables in question. Section~\ref{sec:T_N} is concerned with determining the temperature of the N\'eel transition. Sec.~\ref{sec:thermodynamics} describes the thermodynamics near $T_N$.  Here we discuss the extrapolation of the observables to the TD limit (\ref{subsec:TD-limit_extrapolation}), the determination of entropy (\ref{subsec:entropy}), and thermometry near $T_N$ (\ref{subsec:thermometry}). We summarize the results in Sec.~\ref{sec:conclusions}. The Appendix contains tables of the obtained numerical data for the entropy, energy, double occupancy, and spin-spin correlation functions as a function of temperature.  

\section{Method} \label{sec:method} 

We first rewrite the Hubbard Hamiltonian \eqref{Hubbard} to a form suitable for numerical simulations by mapping the repulsive model \eqref{Hubbard} to an  attractive model by a particle-hole transformation \cite{Micnas1990}. We use the fact that the simple cubic lattice is bipartite and can be split into two interpenetrating sublattices $\mathcal{A}$ and $\mathcal{B}$, so that the hopping term in \eqref{Hubbard} only connects sites belonging to different sublattices. Then we introduce the hole operators for the $\up$-component:
\begin{equation}
a_{\vv{x}\up}^\dagger = \left\{%
\begin{aligned}
c_{\vv{x}\up}\,, &\quad\vv{x}\in\mathcal{A} \\ 
-c_{\vv{x}\up}\,, &\quad\vv{x}\in\mathcal{B} \\ 
\end{aligned}
\right.
\end{equation}
This way, Eq.\ \eqref{Hubbard} becomes at half filling 
\begin{multline}
\widehat{H} = -t\sum_{\langle \vv{x}\vv{y} \rangle \sigma} a^\dagger_{\vv{x}\sigma} a_{\vv{y}\sigma} + \mathrm{h.c.} %
-U\sum_{\vv{x}} m_{\vv{x}\uparrow}   m_{\vv{x}\downarrow} \\%
-\mu'\sum_{\vv{x}\sigma} m_{\vv{x}\sigma} -\frac{U}{2}\Omega \;,
\label{Hubb_attr}
\end{multline}  
where $a_{\vv{x}\down} = c_{\vv{x}\down}$, $m_{\vv{x}\sigma} = a^\dagger_{\vv{x}\sigma} a_{\vv{x}\sigma}$ is the number operator for the attractive model, $\mu' = -U/2$ as appropriate for half filling,  and $\Omega=L^3$ is the total number of sites. 

Since we only consider half filling, $\la m_{\vv{x}\up}\ra = \la m_{\vv{x}\down}\ra = 1/2$, we follow Ref.~\onlinecite{Rubtsov2005} and shift the chemical potential according to $\mu'\to \mu'+\alpha U$\,:
\begin{gather}
\widehat{H} = \widehat{H}_0 + \widehat{H}_1 + \left(\alpha^2 - \frac{1}{2}\right) U\Omega \\
\intertext{with}
\begin{aligned}
\widehat{H}_0  &= -t\sum_{\langle \vv{x}\vv{y} \rangle \sigma} a^\dagger_{\vv{x}\sigma} a_{\vv{y}\sigma} + \mathrm{h.c.} -(\mu'+\alpha U)\sum_{\vv{x}\sigma} m_{\vv{x}\sigma} \\
\widehat{H}_1 &= -U\sum_{\vv{x}} \left( m_{\vv{x}\uparrow} - \alpha\right) \left(  m_{\vv{x}\downarrow} -\alpha \right)
\end{aligned}
\label{Hubb_attr_sym}
\end{gather}
At half filling the choice $\alpha=1/2$ is optimal because it leads to the minimal computational complexity of the simulations (see below), and we use this value of $\alpha$ throughout.

\subsection{Diagrammatic Determinantal Monte Carlo method}  \label{subsec:DDMC}

The DDMC algorithm works with the weak-coupling expansion series for the finite-temperature partition function for the Hubbard model \eqref{Hubb_attr_sym}. The latter reads
\begin{gather}
Z = Z_0\sum_{p=0}^{\infty} U^p \sum_{\vv{x}_1 \dots \vv{x}_p} \int_{0<\tau_1<\dots<\tau_p<\beta}\prod_{j=1}^{p} %
\mathcal{D}(\vv{x}_1\tau_1; \dots; \vv{x}_p\tau_p)\;,
\label{Z_1}
\intertext{where}
\mathcal{D}(\vv{x}_1\tau_1; \dots \vv{x}_p\tau_p) = %
\left\la \prod_{j=1}^{p}\left( m_\up(\vv{x}_j\tau_j) -\alpha \right) \left( m_\down(\vv{x}_j\tau_j)-\alpha \right)%
 \right\ra_0 \;.
\label{Z_2}
\end{gather} 
Here $\beta$ is the inverse temperature, $Z_0 = \Tr \mathcal{T} \exp{(-\beta H_0)}$ is the unperturbed partition function, $\mathcal{T}$ denotes time ordering, and $\la (\dots) \ra_0 = \Tr \left[ \mathcal{T} (\dots)\exp{(-\beta H_0}) \right]/Z_0$ denotes the thermodynamic average with respect to the unperturbed Hamiltonian $H_0$.

Equations \eqref{Z_1}--\eqref{Z_2} generate the standard Feynman diagrams: There are $(p!)^2$ diagrams of order $p$, which can be represented graphically as a collection of $p$ vertices connected by single-particle propagators. Summing over all the interconnections for a fixed vertex configuration
\begin{equation}
\mathcal{S}_p = \{ (\vv{x}_j \tau_j), j=1,\dots,p\}
\label{Sp}
\end{equation}
equation \eqref{Z_2} takes the form
\begin{equation}
\mathcal{D}(\mathcal{S}_p) = | \det\mathrm{A}(\mathcal{S}_p) |^2\;,
\label{Z_det}
\end{equation}
where $\mathrm{A}(\mathcal{S}_p)$ are $p\times p$ matrices with matrix elements given by 
($i,j = 1,\dots,p$)
\begin{equation}
A_{ij}(\mathcal{S}_p) = G^{(0)}( \vv{x}_i-\vv{x}_j, \tau_i-\tau_j ) - \alpha\delta_{ij} 
\label{A_matr_el}
\end{equation}
(since we only consider an unpolarized system, we omit the spin index for $\mathrm{A}$-s and $G^{(0)}$-s), and $G^{(0)}$ being the free-particle Green's functions,
\begin{equation}
G^{(0)}( \vv{x}_i-\vv{x}_j, \tau_i-\tau_j ) = - \la a_{\vv{x}_i}(\tau_i) a^\dagger_{\vv{x}_j}(\tau_j)  \ra_0. \label{G_0_definition}
\end{equation} 
Since $\hat{H_0}$, Eq.~\eqref{Hubb_attr_sym}, is diagonal in momentum space,
\begin{eqnarray}
\widehat{H}_0  = \sum_{ \vv{k} \sigma} [\varepsilon_\vv{k} - \mu' -\alpha U]  a^\dagger_{\vv{k}\sigma} a_{\vv{k}\sigma}, \nonumber \\
\varepsilon_\vv{k} = - 2t\sum_{i=1,2,3} \cos(k_i), \label{H_0_k}
\end{eqnarray}
the free propagators are calculated by the Fourier transform 
\begin{eqnarray}
G^{(0)}( \vv{r}, \tau) = \sum_\vv{k} G^{(0)}( \vv{k}, \tau) e^{-i \vv{k} \vv{r}}, \nonumber \\ 
G^{(0)}( \vv{k}, \tau) = - \frac{e^{-(\varepsilon_\vv{k} -\mu' -\alpha U)\tau}}{[1+e^{-\beta (\varepsilon_\vv{k}-\mu' -\alpha U)}]}, \; \tau>0, \label{G_0_FT} \\
G^{(0)}( \vv{k}, -\tau)=-G^{(0)}( \vv{k}, \beta-\tau), \nonumber
\end{eqnarray}
and tabulated before the start of the simulation.

The series \eqref{Z_1}, \eqref{Z_det}--\eqref{A_matr_el} serves as a basis for a DDMC simulation of the Hubbard model \eqref{Hubb_attr_sym}: We set up a random walk in the space of the vertex configurations $\mathcal{S}_p$, Eq.\ \eqref{Sp}, using the Metropolis algorithm \cite{Metropolis} with the weights proportional to $\mathcal{D}(\mathcal{S}_p)$, Eq.\ \eqref{Z_det}. 
Since the technique itself is detailed elsewhere \cite{Rubtsov2005, DDMC_njp2006}, here we only briefly discuss the specific details of the present implementation. We only stress at this point that since all the terms in the series \eqref{Z_1},\eqref{Z_det}--\eqref{A_matr_el} are positive definite we completely avoid a sign problem. 

The simplest updating strategy for DDMC simulations consists of adding ($p\to p+1$) and removing ($p\to p-1$) interaction vertices at random positions in $\vv{x}$ and $\tau$ to/from a vertex configuration $\mathcal{S}_p$. However, at half filling and with $\alpha=1/2$ the series only contains even order terms, hence we employ rank-2 updates, where $p\to p\pm 2$. Using the Woodbury-type formulas, both rank-1 and rank-2 updates can be performed in $O(p^2)$ operations.
We note that for $\alpha \neq 1/2$ both even and odd terms are present even at half filling. In this sense the choice of $\alpha=1/2$ is optimal.

\subsection{Observables} \label{subsec: observables} 

The general method for calculating observables in the DDMC simulations uses the standard technique of Monte Carlo estimators: for an observable $\mathcal{O}$  we define an estimator $\mathcal{Q}^{(\mathcal{O})}(\mathcal{S}_p)$ such that the average of the latter over the vertex configurations generated by the MC process, $\la \mathcal{Q}^{(\mathcal{O})}(\mathcal{S}_p) \ra_{\scriptsize \mathrm{MC}}$, converges to the thermal average $\la \mathcal{O} \ra$, where 
\begin{equation}
\la \mathcal{O} \ra = Z^{-1} \Tr \left[ \mathcal{O}\, e^{-\beta H} \right] \;.
\end{equation}
Below we explicitly list estimators for useful observables: 

\paragraph{Filling fraction}   

The thermal average of the filling fraction $m_{\sigma}$ for the spin projection $\sigma$ is given by
\begin{equation} 
\la m_{\sigma} \ra = \la a^\dagger_{\vv{x}\sigma} a_{\vv{x}\sigma} \ra 
\label{dens}
\end{equation}
hence the corresponding estimator is\cite{DDMC_njp2006}
\begin{equation}
\mathcal{Q}^{(n_\sigma)}(\mathcal{S}_p) = \alpha + \frac{\det\mathbf{B}(\mathcal{S}_p ; \vv{x}\tau)}{ \det\mathbf{A}(\mathcal{S}_p)} \;,
\label{Q_dens}
\end{equation}
where $\mathbf{A}(\mathcal{S}_p)$ is a $p\times p$ matrix \eqref{A_matr_el} and $\mathbf{B}(\mathcal{S}_p ; \vv{x}\tau)$ is a $(p+1)\times (p+1)$ matrix with an extra row and a column corresponding to the extra creation and annihilation operators in \eqref{dens}. Here $\vv{x}$ and $\tau$ are random positions in space and time, respectively.

Notice that for a half filled model and for $\alpha=1/2$, the MC average of the second term in \eqref{Q_dens} must equal zero. We use this fact to check if a simulation has equilibrated. 

\paragraph{Kinetic energy} 

Calculating the kinetic energy for the model \eqref{Hubb_attr_sym} requires evaluating the average of $\la a^\dagger_{\vv{x}\sigma} a_{\vv{y}\sigma} \ra$, which differs from Eq.\ \eqref{dens} only in that the creation annihilation is shifted in space with respect to the creation operator while in Eq.\ \eqref{dens} both operators reside on the same lattice site. The estimator for the kinetic energy $\la H_0 \ra$ is then\cite{DDMC_njp2006}
\begin{equation}
\mathcal{Q}^{(H_0)}(\mathcal{S}_p) = -2t \frac{\det\mathbf{B}(\mathcal{S}_p ; \vv{x},\vv{y},\tau)}{ \det\mathbf{A}(\mathcal{S}_p)} \times 2\,d\,L^3 \;,
\label{KE_Q}
\end{equation}
where the matrix $\mathbf{B}(\mathcal{S}_p ; \vv{x},\vv{y},\tau)$ differs from $\mathbf{B}(\mathcal{S}_p ; \vv{x}\tau)$ by the fact that the creation operator in \eqref{KE_Q} is shifted in space with respect to the annihilation operator. The extra factor of 2 accounts for the summation over $\sigma=\up,\down$ and $dL^3$  is the number of bonds of a lattice with $L^3$ sites and periodic boundary conditions (PBC).

\paragraph{Interaction energy} 

Since the series \eqref{Z_1} is nothing but an expansion in powers of $\widehat{H}_1$, the corresponding estimator is readily obtained by a standard trick of considering the Hamiltonian $H_0 + \lambda H_1$ and differentiating with respect to $\lambda$ . The result is\cite{DDMC_njp2006}
\begin{equation}
\mathcal{Q}^{(H_1)}(\mathcal{S}_p) = -\frac{p}{\beta}
\label{PE_Q}
\end{equation}

\paragraph{AFM structure factor} 

Calculating the AFM structure factor \eqref{AFM} in the particle-hole transformed model \eqref{Hubb_attr_sym} requires calculating two independent equal-time density-density correlation functions 
\begin{gather} 
g_{\down\down}(\vv{x}-\vv{y}) = \la m_{\vv{x}\down}\, m_{\vv{y}\down}  \ra \;, \label{guu}%
\\
g_{\up\down}(\vv{x}-\vv{y}) = \la m_{\vv{x}\up}\, m_{\vv{y}\down}  \ra \;,
\label{gud}
\end{gather}

The estimator for the equal spin density correlation function $g_{\up\up}$ for $\vv{x}\neq\vv{y} $ is given by 
\begin{equation}
\mathcal{Q}^{(g_{\down\down})}(\mathcal{S}_p) = \alpha^2 + \frac{ \det\mathbf{B}_2(\mathcal{S}_p ; \vv{x},\vv{y},\tau) }{ \det\mathbf{A}(\mathcal{S}_p) }
\label{Q_gud}
\end{equation}
where $\mathbf{B}_2(\mathcal{S}_p ; \vv{x},\vv{y},\tau)$ is a $(p+2)\times(p+2)$ matrix with two extra rows and columns corresponding to the extra creation and annihilation operators in Eq.\ \eqref{guu}. For $\vv{x}=\vv{y}$,  $g_{\up\up}(\vv{x}=\vv{y}) = \la m_{\vv{x}\up}^2 \ra$ which equals $1/2$ for a half-filled model.

To build the estimator for Eq.\ \eqref{gud} we proceed similarly to \eqref{PE_Q}. The resulting estimator for \eqref{gud} is

\begin{equation}
\mathcal{Q}^{(g_{\up\down})}(\mathcal{S}_p) = \alpha^2 + \frac{p}{\beta U \Omega} \frac{ \det\overline{\mathbf{B}}(\mathcal{S}_p ; \vv{y},\tau) }{ \det\mathbf{A}(\mathcal{S}_p) }
\label{Q_guu}
\end{equation}
where $\overline{\mathbf{B}}(\mathcal{S}_p ; \vv{y},\tau)$ is a $p\times p$ matrix constructed by selecting a random vertex, $(\vv{x}\tau)$, from a configuration $\mathcal{S}_p$ and moving the corresponding row of the matrix $\mathbf{A}(\mathcal{S}_p)$ to $(\vv{y}\tau)$ while leaving the corresponding column at $(\vv{x}\tau)$\,.

\section{Critical temperature} \label{sec:T_N} 

In the paramagnetic phase, $T>T_N$, $S(\vv{Q})$ scales to zero exponentially as $L\to\infty$. In the AFM phase, on the other hand, $S(\vv{Q})/L^{3}\to m^2$ as $L\to\infty$, where $m$ is the sublattice magnetization. Right at the critical temperature $S(\vv{Q}) \propto 1/L^{-2+\eta}$, where $\eta$ is the anomalous dimension. 

In order to locate the transition temperature we thus use the standard finite-size scaling (FSS) ansatz \cite{Binder1981}
\begin{equation}
S(\vv{Q}) L^{-2+\eta} = f(L/\xi) (1+c L^{-\omega} + \dots)\;, 
\label{NMSF}
\end{equation}
where $\xi$ is the correlation length which diverges at the transition as $\xi\propto |T-T_N|^{-\nu}$, $f(x)$ is a real-valued function which tends to a finite constant as $x\to 0$, and the corrections in brackets arise from the leading irrelevant operators (dots represent the higher-order corrections). Here the exponent $\omega$ is universal, but the amplitude $c$ is not. In accordance with the 3D Heisenberg universality class, we take $\eta\approx 0.037$, $\nu\approx 0.71$ and $\omega\approx 0.8$ \cite{Hasenbusch2002}.  

\begin{figure}[!h]
\includegraphics[width = 0.98\columnwidth,keepaspectratio=true]{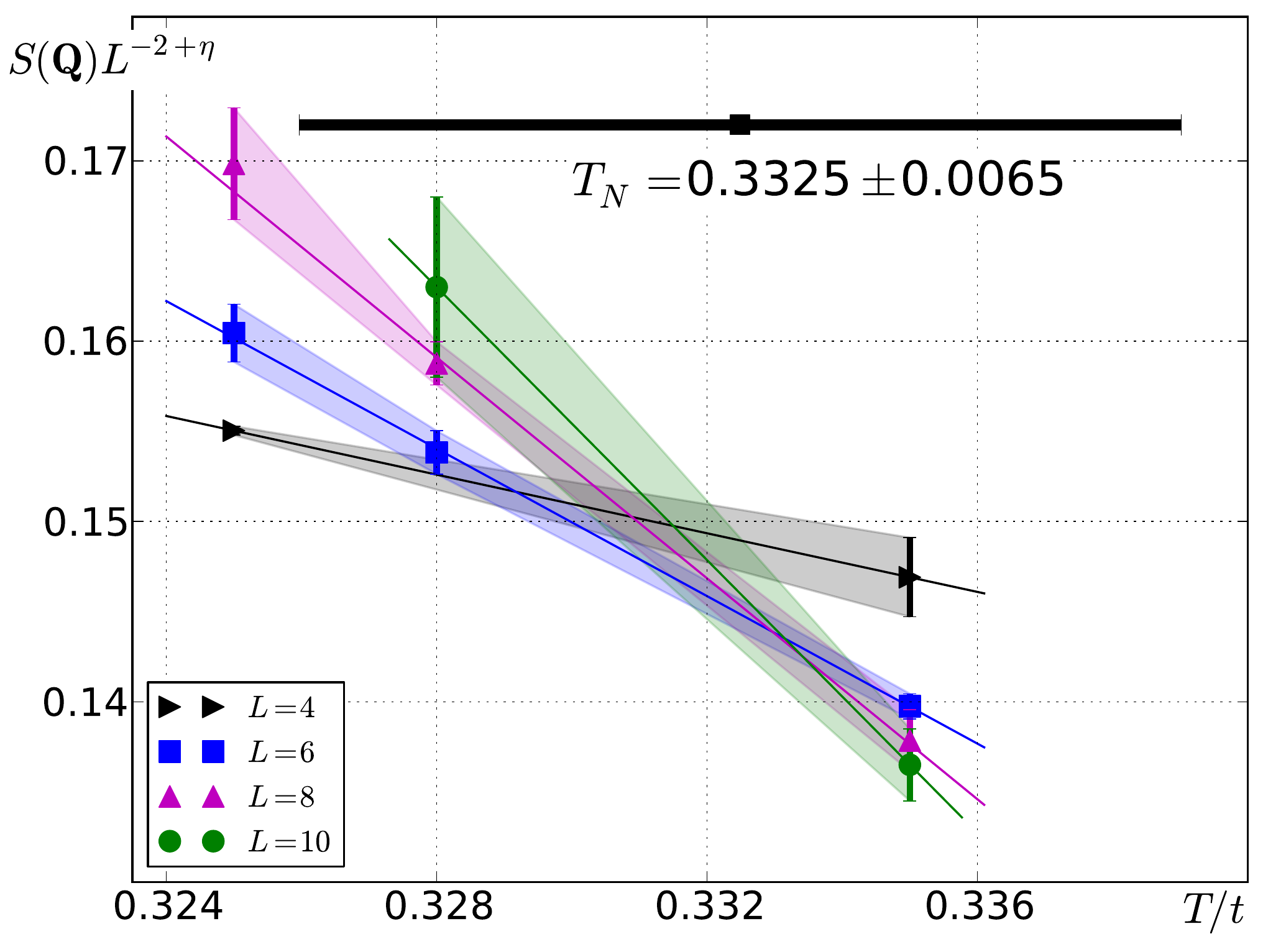}
\caption{(Color online.) Finite size scaling for $T_N$ at $U=8t$ by Binder crossings analysis. 
Points are Monte Carlo results, lines are linear fits. The uncertainty for the N\'eel temperature is estimated conservatively by varying the Monte Carlo points within their respective errorbars.}
\label{fig:Tc_U8}
\end{figure}

The basic idea of using Eq.\ \eqref{NMSF} for the FSS is as follows: if the corrections-to-scaling (the 2nd term in brackets in Eq.\ \eqref{NMSF}) were not present, $S(\vv{Q})L^{-2+\eta}$ would be scale independent at the transition point, so that performing the simulations at a series of system sizes $L_1 > L_2 > \dots$ and
plotting $S(\vv{Q})L^{-2+\eta}$ versus temperature, one would observe that all the curves intersect at the same point, $T=T_N$. This is what we indeed observe for (relatively) large values of $U/t$ at $L \geq 6$: for $U\geq 6t$ our MC results are consistent with $c=0$ in Eq.\ \eqref{NMSF} within statistical errors, see Fig.\ \ref{fig:Tc_U8}. The data for the smallest systems of size $L=4$ systematically deviate from the scaling described by Eq.\ \eqref{NMSF} for all values of $U$ considered here (see also below), and hence they are omitted from the scaling analysis. 

\begin{figure}[!h]
\includegraphics[width = 0.98\columnwidth,keepaspectratio=true]{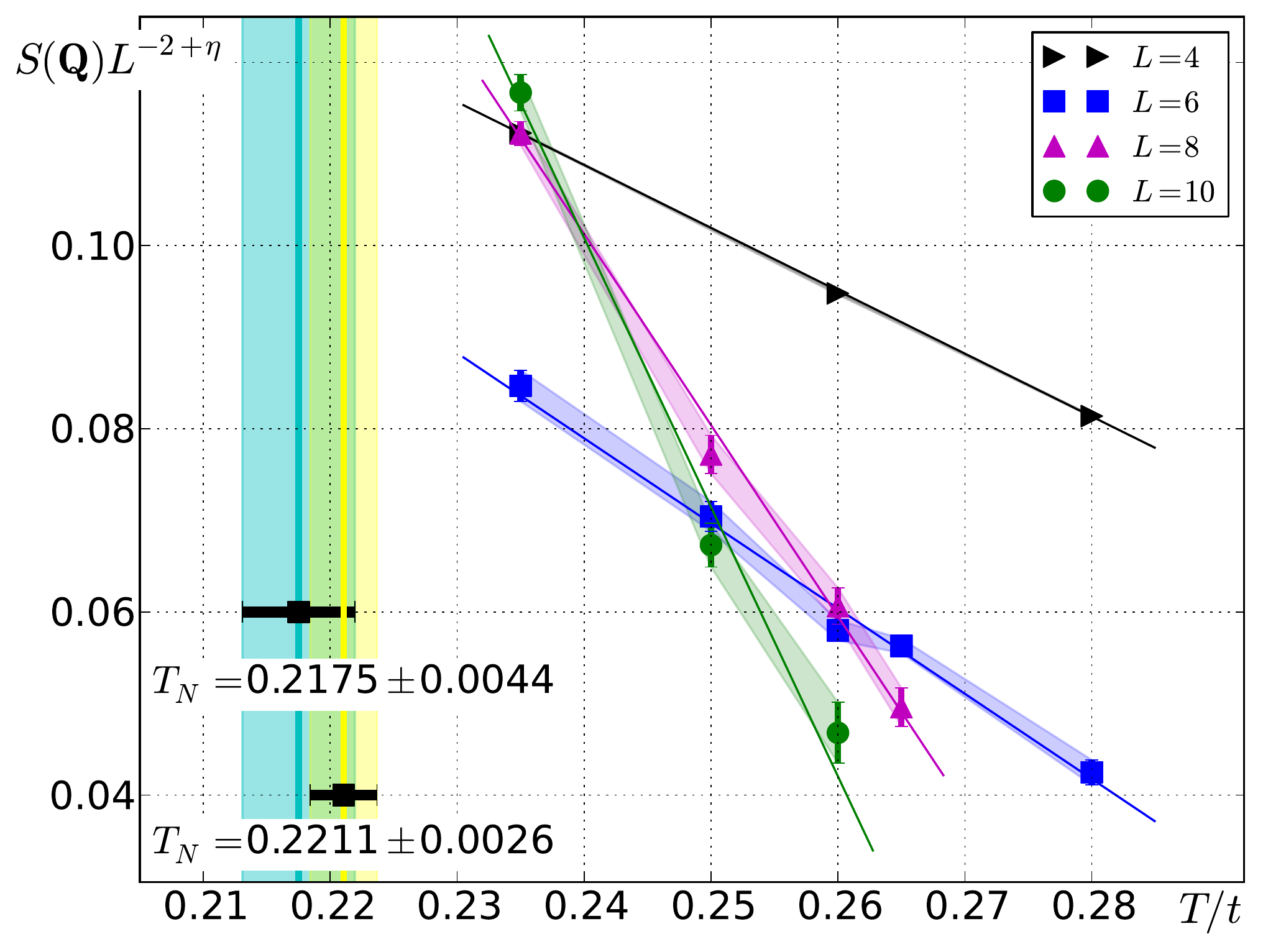} 
\caption{(Color online.) Finite size scaling for 
$T_N$ at $U=5t$ by Binder crossings analysis. Points with errorbars are Monte Carlo results, solid lines are linear fits. FSS procedures based on Eqs.\ \eqref{binder_cross}--\eqref{binder_cross_rhs} and \eqref{gw_way} result in the estimates $T_N/t=0.2175(44)$ and 
$T_N/t=0.2211(26)$, 
respectively. See also Fig.~\ref{fig:OmegaU5}. 
}
\label{fig:Tc_U5}
\end{figure}

\begin{figure}[!h]
\includegraphics[width = 0.98\columnwidth,keepaspectratio=true]{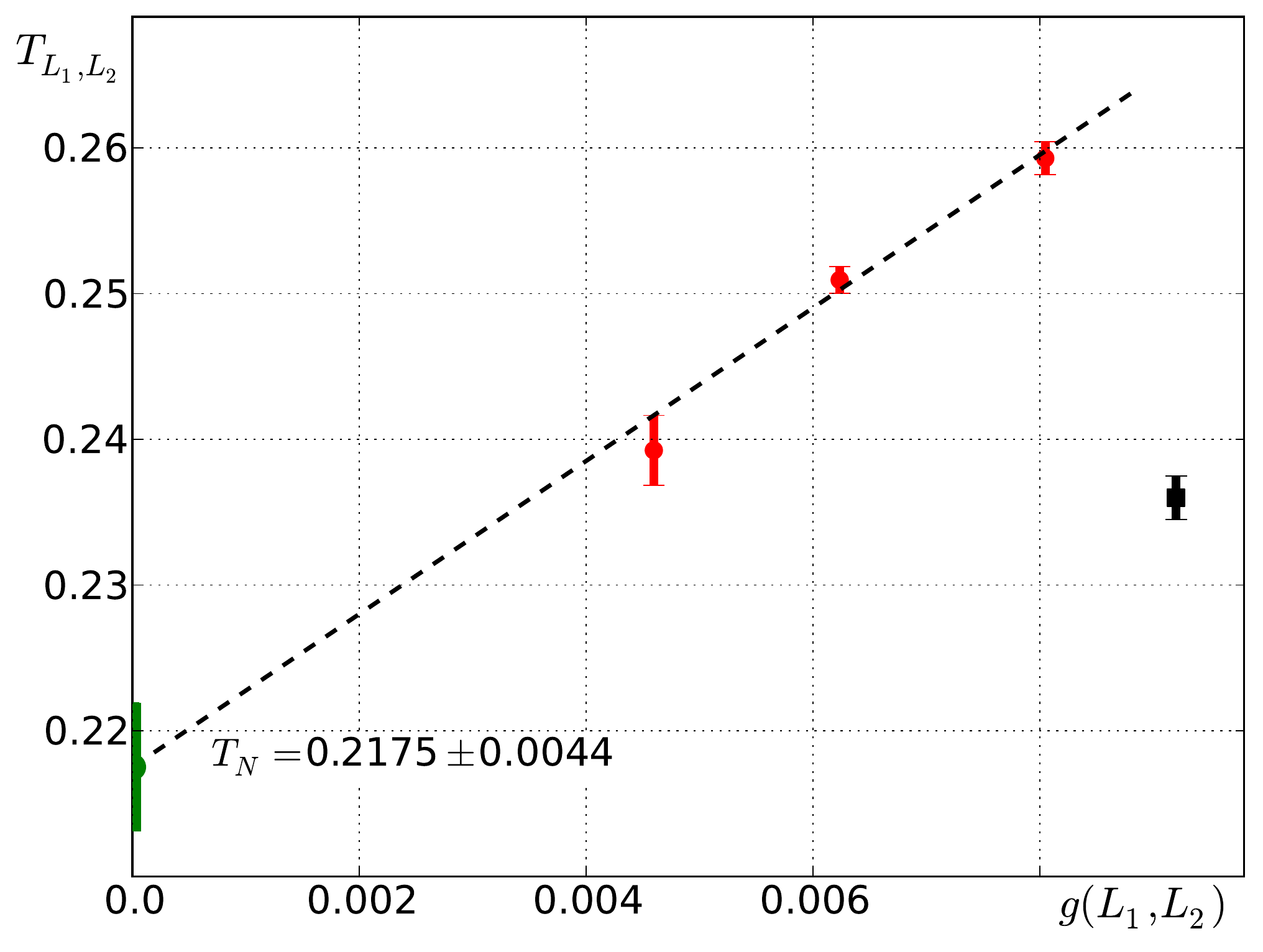} 
\caption{(Color online.) Scaling---according to Eqs.\ \eqref{binder_cross}--\eqref{binder_cross_rhs}---of estimates of the critical temperature at $U=5t$ obtained from Binder crossings between lines for different system sizes in Fig.~\ref{fig:Tc_U5}. See text for discussion. The square corresponds to the crossing point between lines for $L=4$ and $L=10$, which substantially deviates from the linear scaling exhibited by all the crossings for $L > 4$ (circles), demonstrating that the $L=4$ system is too small to be consistent with the critical scaling described by Eq.\ \eqref{NMSF}. Correspondingly, all the other crossings with the $L=4$ line are omitted from the figure.}
\label{fig:OmegaU5}
\end{figure}

We find that the corrections-to-scaling become more pronounced with decreasing $U/t$: For $U=5t$ we find a clear evidence of the shift of the pairwise crossings towards lower temperatures, see Fig.\ \ref{fig:Tc_U5}. Since simulating larger system sizes is not an option, we employ Eq.\ \eqref{NMSF} including corrections to scaling. The most straightforward way is to follow the evolution of the pairwise crossings with the system size. Indeed, expanding Eq.\ \eqref{NMSF} around the crossing of $S(\vv{Q}) L^{-2+\eta}$ at system sizes $L=L_1$ and $L=L_2$ up to the terms linear in $T-T_N$ we find (cf Ref.~\onlinecite{DDMC_njp2006}):
\begin{align}
T_{L_1, L_2} -T_N &= \mathrm{const} \times g(L_1,L_2) \;,
\label{binder_cross}
\intertext{where}
g(L_1,L_2) &= \frac{1}{L_2^{1/\nu+\omega}} \frac{(L_2/L_1)^\omega-1}{1-(L_1/L_2)^{1/\nu}} \;.
\label{binder_cross_rhs}
\end{align}
We perform the linear fit of the series of crossings $T_{L_1,L_2}$ versus $g(L_1,L_2)$. Then the intercept of the best-fit line yields the N\'{e}el temperature. This procedure is illustrated in Fig.\ \ref{fig:OmegaU5}. It is clear from Figs.~\ref{fig:Tc_U8}--\ref{fig:OmegaU5} that $L=4$ does not follow the scaling \eqref{binder_cross}--\eqref{binder_cross_rhs}. We attribute it to the effect of the higher-order terms neglected in \eqref{NMSF}, and only use $L>4$ in the fitting procedure \eqref{binder_cross}--\eqref{binder_cross_rhs}.


We note at this point that the methodology based on Binder crossings has a build-in self-consistency check: if, in fact, the criticality were not in the Heisenberg universality class, the curves for the magnetic structure factor, rescaled via Eq.\ \eqref{NMSF} would have no reason to cross at a unique point, and thus the whole procedure of \eqref{NMSF}--\eqref{binder_cross_rhs} would break down. 

An equivalent procedure has been suggested in Ref.~\onlinecite{Goulko2010}. Again, one expands Eq.\ \eqref{NMSF} up to the linear order in $T-T_N$, which leads to
\begin{equation}
S(\vv{Q})L^{-2+\eta} = \left( a_0 + a_1(T-T_N) L^{1/\nu} \right) \left( 1+ c L^{-\omega} \right) \;,
\label{gw_way}
\end{equation}
which is then used as a four-parameter ansatz for a single nonlinear fit. \textit{A priori}, fitting procedures based on 
\eqref{binder_cross} and \eqref{gw_way} are equivalent and indeed produce consistent results, as illustrated in Fig. \ref{fig:Tc_U5}. We stress at this point that using \eqref{gw_way} requires judicious choice of the temperature range for fitting: including Monte Carlo points at too high temperatures and/or too small system sizes tends to significantly skew the fit results. In the following we therefore quote the $T_N$-s obtained using Eq.\ \eqref{binder_cross}. 
 
\begin{figure}[!h]
\includegraphics[width = 0.98\columnwidth,keepaspectratio=true]{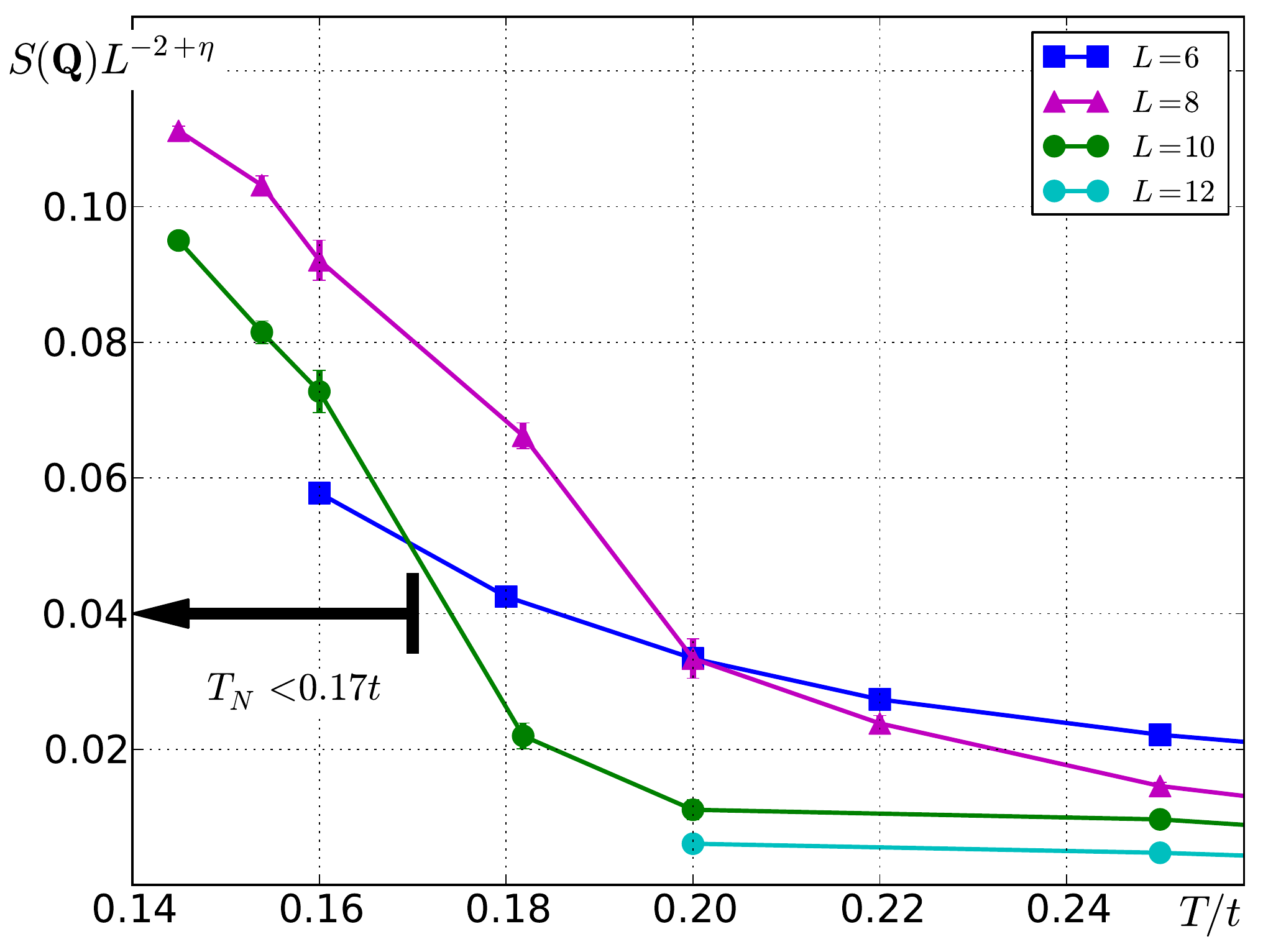} 
\caption{(Color online.) Finite size scaling for $T_N$ at $U=4t$ by Binder crossings analysis. In this case we not able to reliably extract the N\'eel temperature and can only provide an upper limit, $T_N < 0.17 t$. Notice that the crossing of $L=6$ and $L=8$ is clearly outside of the range of applicability of either \eqref{binder_cross} or \eqref{gw_way}.  }
\label{fig:Tc_U4}
\end{figure}

For $U=4t$, we find the corrections-to-scaling to be larger than those for $U=5t$, see Fig.\ \ref{fig:Tc_U4}. In fact, with the accessible systems sizes we are only able to put an upper limit on the N\'eel temperature, $T_N < 0.17 t$. From Fig.\ \ref{fig:Tc_U4} it is clear that $L=6$ and possibly even $L=8$ are simply too small and need to be discarded from the finite-size scaling analysis. 

\begin{figure}[!h]
\includegraphics[width = 0.98\columnwidth,keepaspectratio=true]{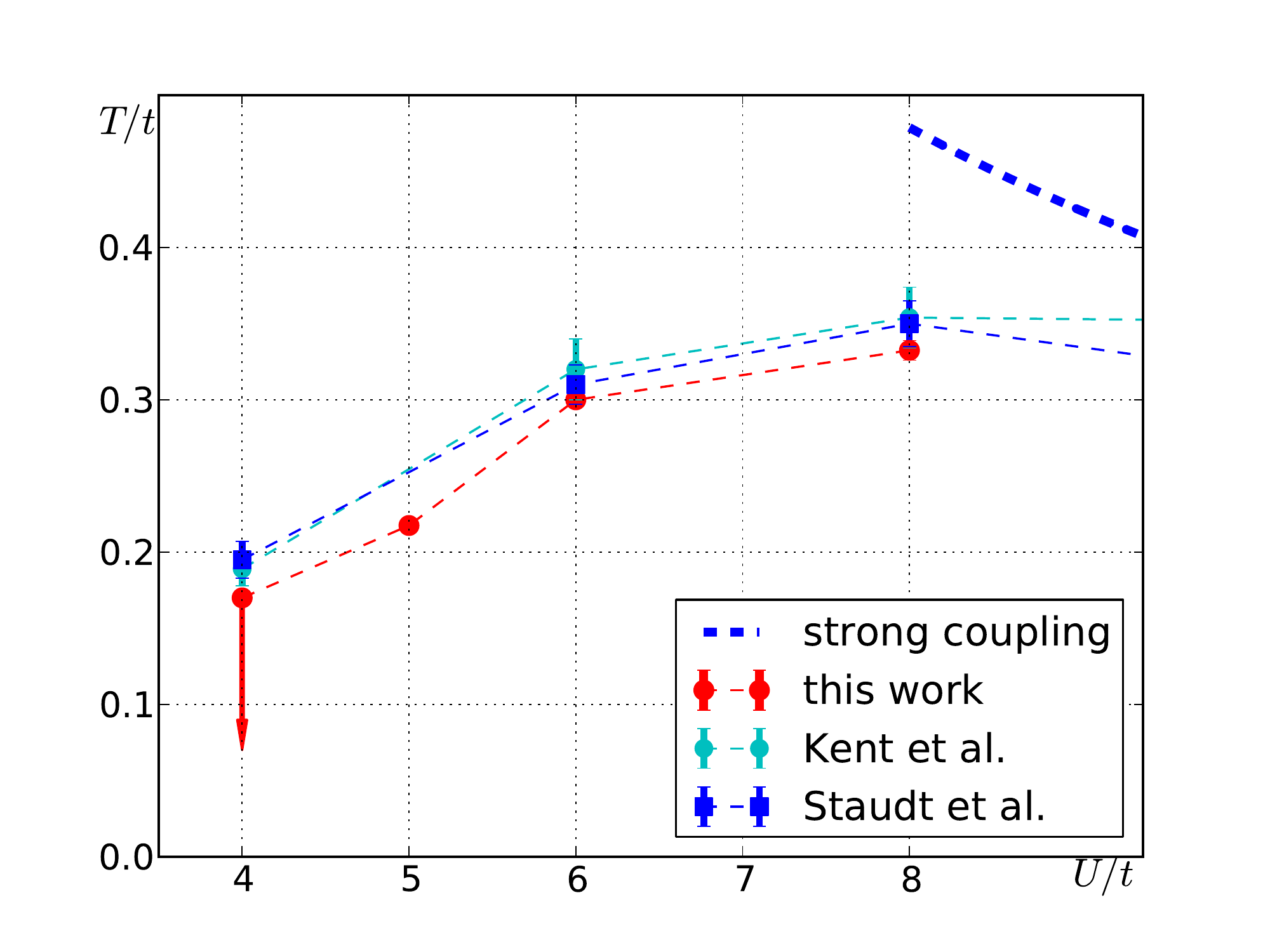}
\caption{(Color online.) Comparison of estimates for $T_N$ by different unbiased approaches. Also shown is the strong-coupling limiting behavior, $T_N=3.83 t^2/U$.  See text for discussion. }
\label{fig:TNs}
\end{figure}

Our results for the dependence of the N\'eel temperature on $U$ are summarized in Table\ \ref{table:TNs} and Fig.\ \ref{fig:TNs}. It is instructive to compare our estimates to the previous unbiased calculations from the literature. While for $U/t=6$ and $8$ our estimates agree with and are more accurate than previous estimates from QMC \cite{Staudt2000} and DCA \cite{Kent2006}. For smaller values of $U/t$ our estimates are systematically lower. The discrepancy can be traced back to the FSS procedure which includes corrections-to-scaling, Eq.\ \eqref{NMSF}: if we were to discard the corrections and identified the Binder crossings of $L=6$ and $L=8$ with the N\'eel temperature, such estimates would have agreed with Refs. \onlinecite{Staudt2000, Kent2006}.  We therefore conclude that the estimates of $T_N$ presented here are more accurate than results reported to date.

\begin{table}
\begin{tabular}{|c|c|c|c|}
\hline
\hline
$U/t$ & $T_N/t$ & $S_N$ \\
\hline
$4$  & $<0.17$  & $<0.17$ \\
$5$  & $0.2175(44)$ &  $0.135(25)$ \\ 
$6$  & $0.300(5)$   &  $0.305(35)$  \\
$8$  & $0.3325(65)$ &  $0.33(3)$ \\
\hline
\hline
\end{tabular}
\caption{N\'eel temperatures and entropies. See text for discussion.}
\label{table:TNs}
\end{table}

\section{Thermodynamics} \label{sec:thermodynamics} 

\subsection{Extrapolation to the thermodynamic limit}  \label{subsec:TD-limit_extrapolation} 

\begin{figure}
\includegraphics[width = 0.98\columnwidth,keepaspectratio=true]{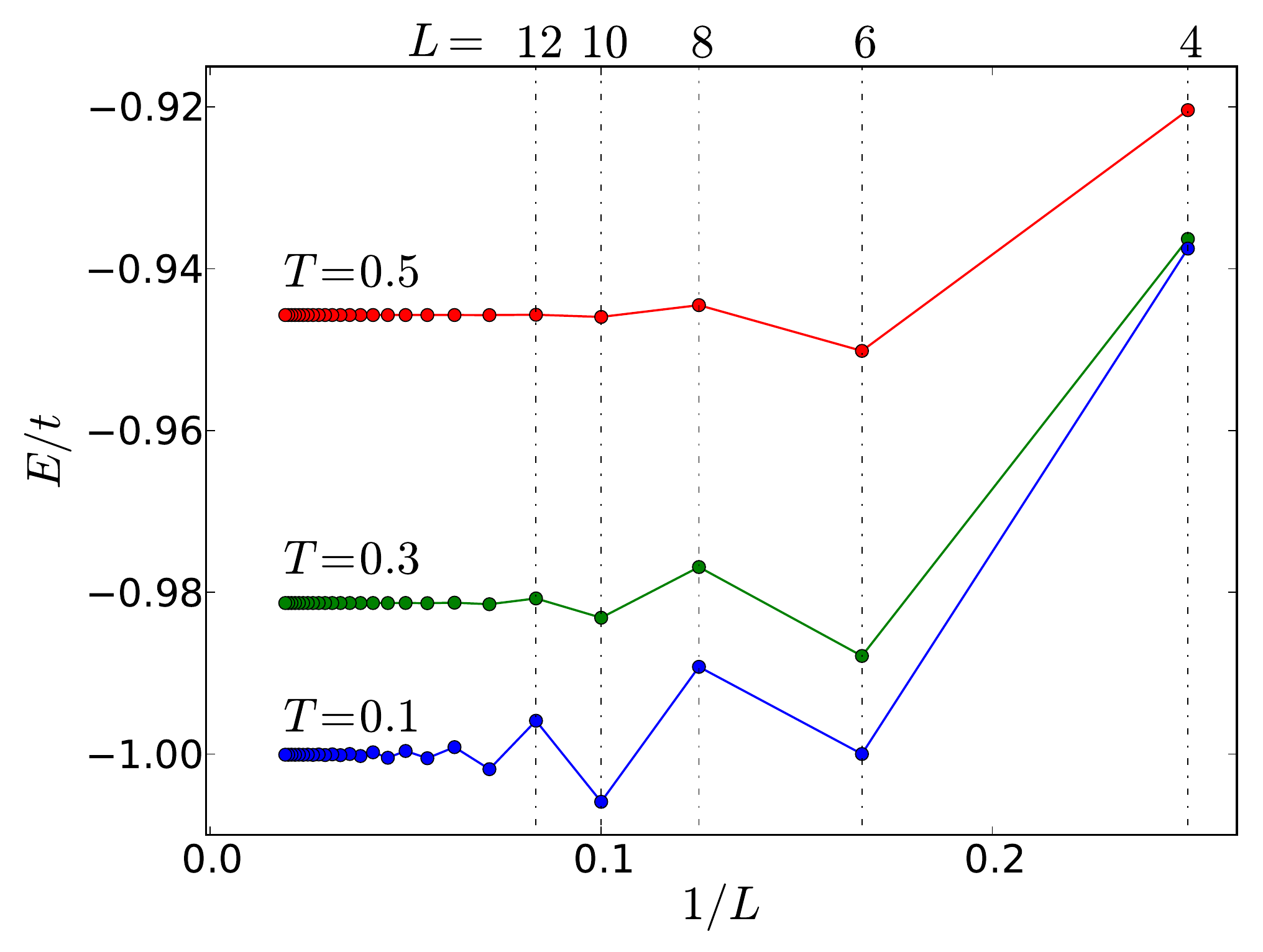}
\caption{(Color online.) Dependence of the energy per particle of non-interacting fermions at half-filling (described by Eq.\eqref{Hubbard} with $U=0, \mu=0$ ) on the inverse of the linear system size $ L $.}
\label{fig:free_fermions}
\end{figure}

The dependence of local observables on the size $ L $ of the system with PBC is complicated \cite{Marsiglio1999} by oscillations between the results for even and odd values of $L/2$.
The issue is illustrated in Fig.~\ref{fig:free_fermions}, where the energy per particle of the half-filled non-interacting system ($U=0, \mu=0$) with PBC is plotted versus $L^{-1}$ up to a large system size ($L=52$) for different temperatures. The TD-limit value is approached from above by the data for even $ L/2 $ and from below by those with $ L/2 $ odd. These are the well-known ``shell'' oscillations \cite{twisted_bc} caused by whether or not the spectrum of the finite system has states $\nu$ with the energy $E_{\nu}$ within a range much less than $T$ from the Fermi level, $|E_{\nu}-\mu| \ll T$. In the example of Fig.~\ref{fig:free_fermions}, the states are classified by the momenta $ k=(k_1, k_2, k_3)$, $k_i=2\pi n_i/L$ with integers $n_i$ taking the values $n_i=- L/2, \ldots, -1, 0, 1, \ldots, L/2-1$. When $\{n_i\}=L/4$, which is only possible if $L/2$ is even, the state $k=(\pi/2, \pi/2, \pi/2)$ is exactly at the Fermi level; it's occupation is $1/2$ (``open shell''), but it gives no contribution to the average energy. Hence, if $ L $ is not large enough so that the spacing between the levels is larger than $ T $, the average energy per particle of the system with a closed shell ($L/2$ odd) is systematically lower than that of the system with an open shell ($L/2$ even) due to the difference of the number of states below the Fermi level, $ L^3/2$ and $ L^3/2-1$ correspondingly. However, for any given temperature $T$ there is a system size $L_*=L_*(T)$ such that for $L>L_*$ the number of states with the energies $|E_{\nu}-\mu| < T$ becomes large removing the distinction between even and odd $L/2$. In the free-particle case of  Fig.~\ref{fig:free_fermions}, the convergence to the TD limit at $L>L_*$ is extremely fast (exponential) with $L_*(T=0.5) \sim 10$, $L_*(T=0.3) \sim 16$, and $L_*(T=0.1) \sim 40$. 

\begin{figure}
\includegraphics[width = 0.98\columnwidth,keepaspectratio=true]{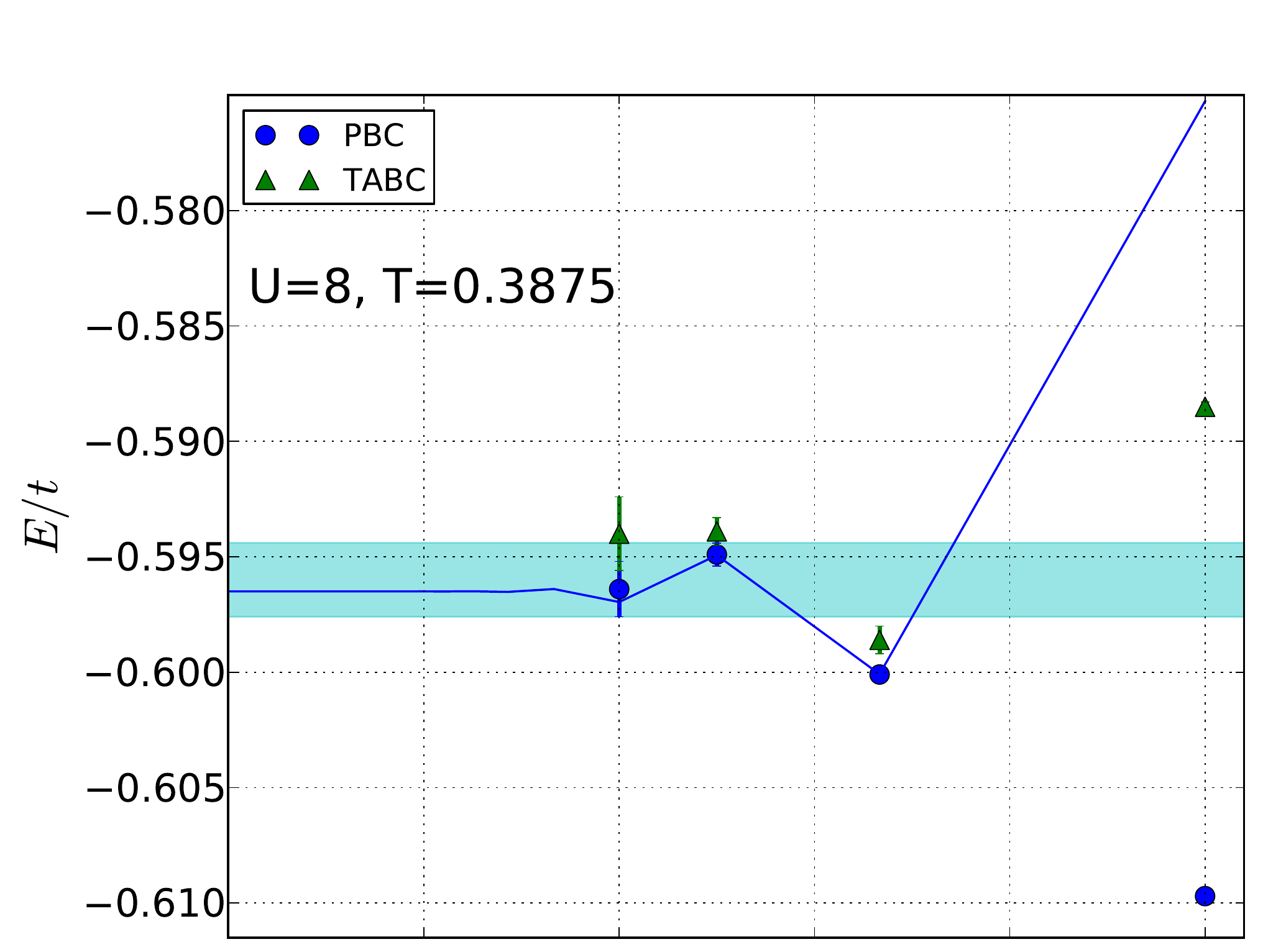}
\includegraphics[width = 0.98\columnwidth,keepaspectratio=true]{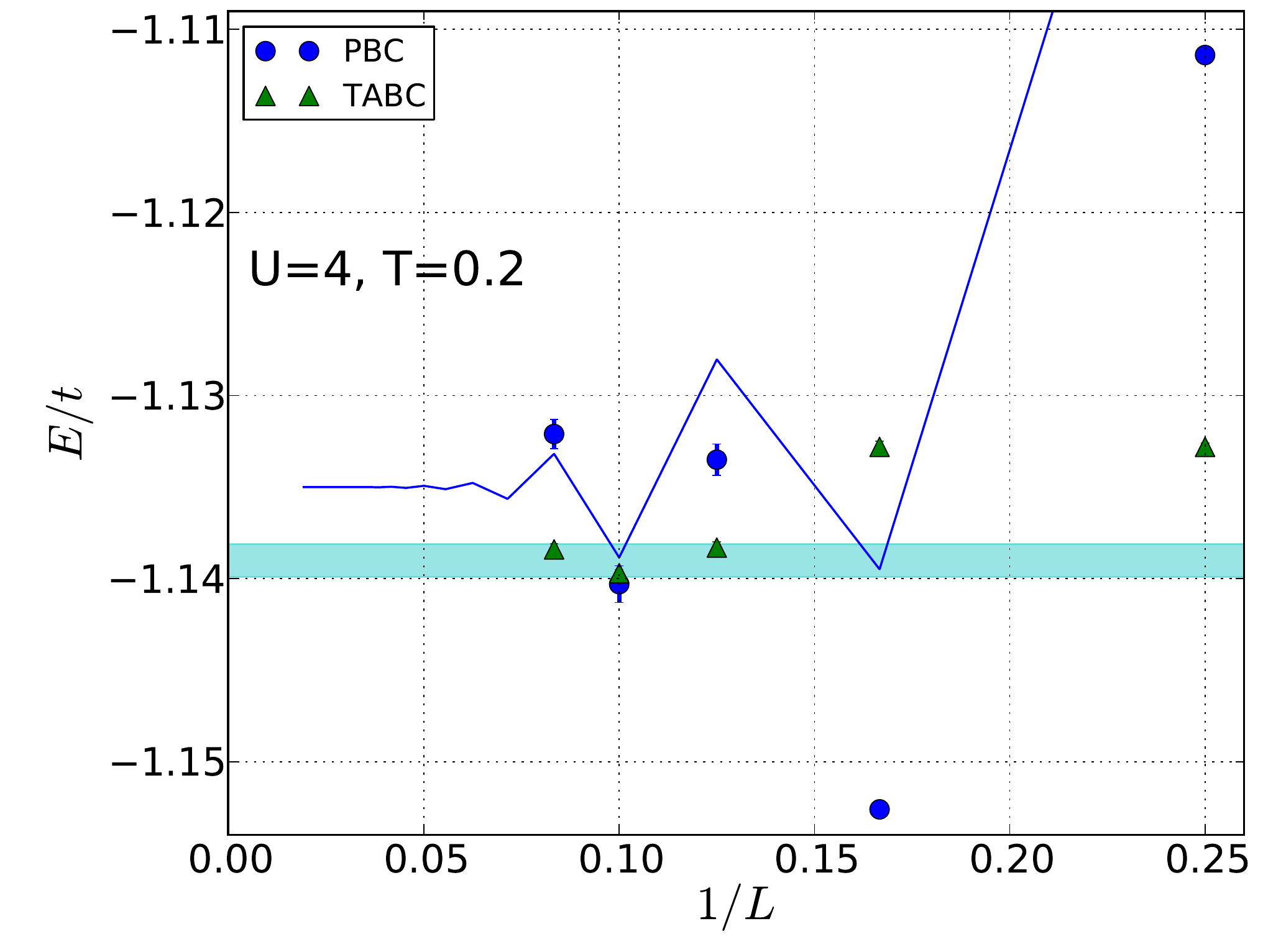}
\caption{(Color online.) Example of the dependence of energy on the inverse of the linear system size $L$ obtained with the periodic boundary conditions (PBC, circles) and using the twist-averaged boundary conditions (TABC, triangles) for $U=8$, $T=0.3875$ and $U=4$, $T=0.2$. The solid line is an extrapolation using the formula $E(L)=E'(\infty) + C [E_0(L)-E_0(\infty) ]$ (see text). For $U=8$, $T=0.3875$ the parameters are $E'(\infty)=-0.5965$, $C=0.6$, while for $U=4$, $T=0.2$, $E'(\infty)=-1.135$, $C=0.95$. The claimed thermodynamic-limit results---$E(\infty)=-0.5960(16)$ for $U=8$, $T=0.3875$ (using the PBC data) and $E(\infty)=-1.1390(9)$ for $U=4$, $T=0.2$ (using the TABC data)---are shown by the horizontal bands. More generally, we use PBC data at $U \geq 6$ and TABC data at $U \leq 5$ to obtain the thermodynamic-limit values, as explained in Subsection.~\ref{subsec:TD-limit_extrapolation}. }
\label{fig:fss1}
\end{figure}

At the finite values of $U$ studied here, the paramagnetic phase should be described by a Fermi liquid  in the limit of $T \ll E_F$. In this regime, the total energy is a functional of occupation numbers of non-interacting quasiparticles. Therefore, the system-size dependence of energy is expected to be proportional to that of the non-interacting system, at least for large enough systems \cite{Ceperley_Alder}. This implies a TD-limit extrapolation in the form $E(L)=E(\infty) + C [E_0(L)-E_0(\infty) ] + g(L)$ suggested in Ref.~\onlinecite{Ceperley_Alder}, where $C$ is a constant, $E_0(L)$ is the energy of the corresponding non-interacting system of size $L$, and $g(L)$ is an unknown in our case function. One can expect that $|g(L)| \ll |C [E_0(L)-E_0(\infty)]|$ for sufficiently large $L$. Given that our simulations are limited to system sizes of up to $L \sim 10$, the validity of this condition is not guaranteed \textit{a priori}.  An example of such an extrapolation with $g(L)=0$ for two typical sets of parameters---$U=8$, $T=0.3875$ and $U=4$, $T=0.2$---is shown in Fig.~\ref{fig:fss1}. The figure suggests that the additional corrections given by $g(L)$ should be small for $L \geq 6$ at large $U$, whereas they are appreciable for most of the accessible system sizes at smaller $U$. We claim the TD-limit value $E(L\to\infty)$ and its error bar $\Delta E(L\to\infty)$ conservatively as the span between the values at the two largest accessible system sizes including their statistical error bars (depicted by the horizontal band in the upper panel of Fig.~\ref{fig:fss1}): 
\begin{multline}
E[L\to\infty] \approx \\ \Big[\mathrm{min}\big(E[L_\mathrm{max}]-\Delta E[L_\mathrm{max}],  E[L_\mathrm{max}-2]-\Delta E[L_\mathrm{max}-2] \big) +\\
\mathrm{max}\big( E[L_\mathrm{max}]+\Delta E[L_\mathrm{max}],  E[L_\mathrm{max}-2]+\Delta E[L_\mathrm{max}-2] \big) \Big]/2 \,,
\label{sheer_horror}
\end{multline}
and
\begin{multline}
\Delta E[L\to\infty] \approx |E[L_\mathrm{max}]-E[L_\mathrm{max}-2]|/2 \\
+ \Delta E[L_\mathrm{max}] + \Delta E[L_\mathrm{max}-2]\,, \label{still_horror}
\end{multline}
and similarly for other local observables. 

As a consistency check for our TD-limit results (at $U \geq 6$) as well as to improve convergence to the TD limit (at $U \leq 5$), we employ two other simulation setups, which exhibit different system-size dependences, which we detail below. 

\subsubsection{Twist-averaged boundary conditions.}

Averaging over twisted boundary conditions was found in Refs.~\onlinecite{Gross, twisted_bc} to produce exact results for the non-interacting system in the grand canonical ensemble and to substantially suppress the system-size dependence for interacting systems. In this approach one introduces a finite phase that particles acquire when they wrap around the periodic boundaries,
\begin{equation}
|\mathbf{r}_1+L \mathbf{e}_i, \mathbf{r}_2, \ldots \rangle = e^{i\Theta_i} |\mathbf{r}_1, \mathbf{r}_2, \ldots \rangle, \;\;\; i=1,2,3, \label{twisted_bc}
\end{equation} 
where $\mathbf{e}_i$ is the unit vector in the direction $i$ and $-\pi < \Theta_i \leq \pi $; Eq.~\eqref{twisted_bc} with $\Theta_i=0$ corresponds to the standard PBC.  Then an observable $A(L)$ is obtained by means of the integration 
\begin{equation}
A(L) =\frac{1}{(2\pi)^3} \int_{-{\bf Q}}^{{\bf Q}} A_\mathbf{\Theta}(L) d\mathbf{\Theta}, \label{observable_with_tbc} 
\end{equation}
where $A_\mathbf{\Theta}(L)$ is a result of the simulation with a fixed value of $\mathbf{\Theta}=\{ \Theta_i\}$ and system size $L$. Thereby, the possible momentum values are forced to span the whole Brillouin zone. The non-interacting propagators $ G^{(0)} $ satisfying the condition \eqref{twisted_bc} are obtained by substituting $\vv{k} \to \vv{k} +\mathbf{\Theta}/L$ in Eq.~\eqref{G_0_FT}. In practice, we perform numerical integration on a mesh of $64$ $ \mathbf{\Theta} $ points, estimating the systematic error of integration to be smaller than the statistical error of $A(L)$ coming from sampling each $A_\mathbf{\Theta}(L)$ by Monte Carlo. 

The results of the calculation of energy with the twist-averaged boundary conditions (TABC) are compared to those for the PBC in Fig.~\ref{fig:fss1}. 

For $U \leq 5$ we find the TABC to substantially reduce the finite size corrections, as exemplified by the lower panel of Fig.~\ref{fig:fss1} showing the typical comparison data for $U=4$. Correspondingly, at $U \leq 5$ we base our $L\to \infty$ extrapolation on the TABC data and use Eqs.~\eqref{sheer_horror},~\eqref{still_horror} with $E[L]$ and $\Delta E[L]$ being the finite-size value and its error bar obtained with TABC to claim the TD-limit extrapolated results (exemplified by the horizontal band in the lower panel of Fig.~\ref{fig:fss1}). 

For larger values of $U/t$, averaging over the twists still reduces the finite size corrections somewhat, but the net improvement of TABC over PBC is smaller (for a typical example see the upper panel of \ref{fig:fss1}---notice that for $U=8$ data for $L=8$ and $L=10$ appear to be converged within their error bars to a value consistent with the PBC extrapolated value)

Overall, we find that for $U/t \geq 6$ the use of TABC does not lead to a significant improvement of the convergence to the TD limit, and we thus use  PBC data and Eqs.~\eqref{sheer_horror},~\eqref{still_horror} in this range of interactions. We follow the same protocol to obtain other observables.


\subsubsection{$L\to\infty$ free propagators.}

\begin{figure}
\includegraphics[width = 0.98\columnwidth,keepaspectratio=true]{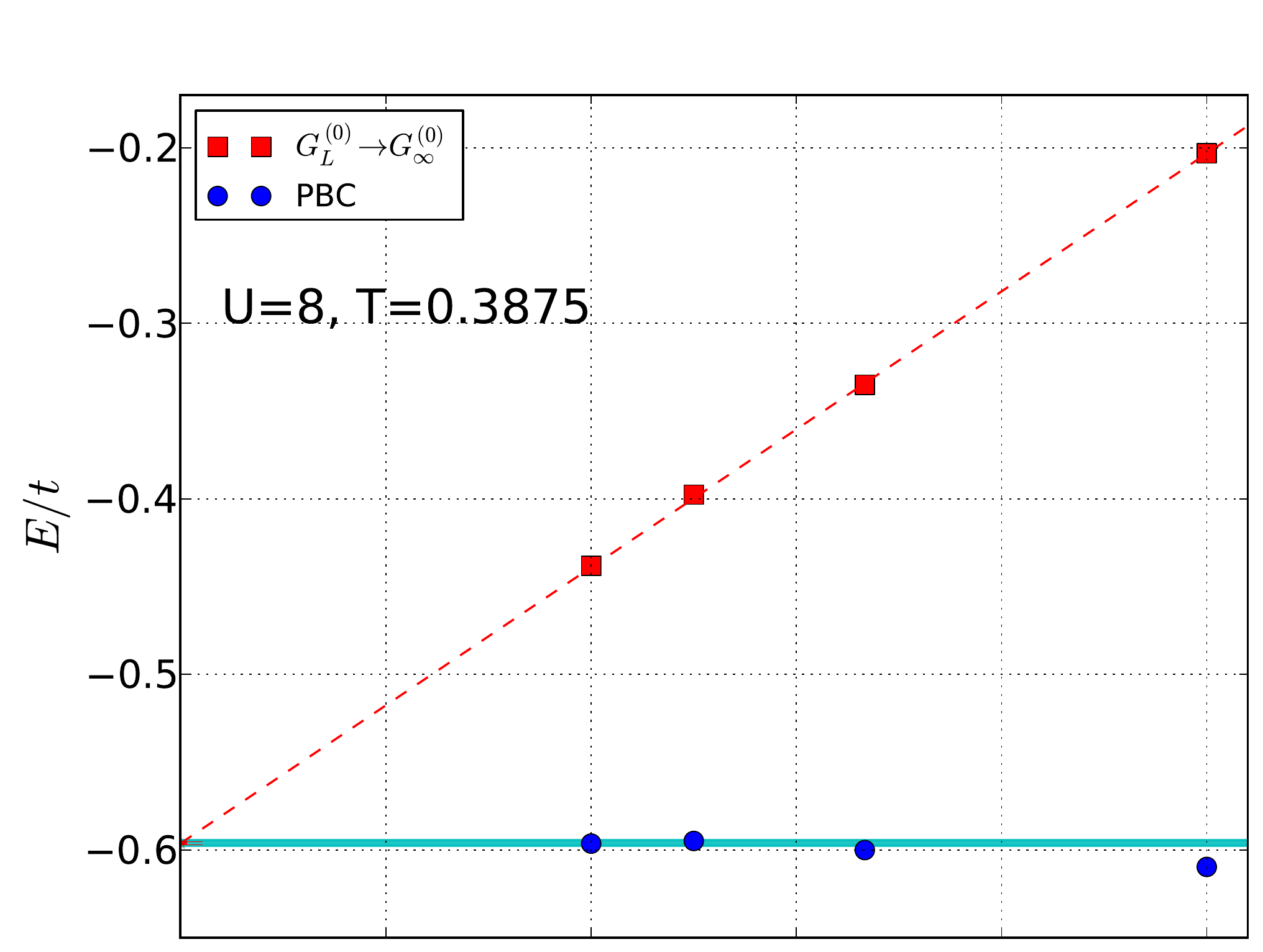}
\includegraphics[width = 0.98\columnwidth,keepaspectratio=true]{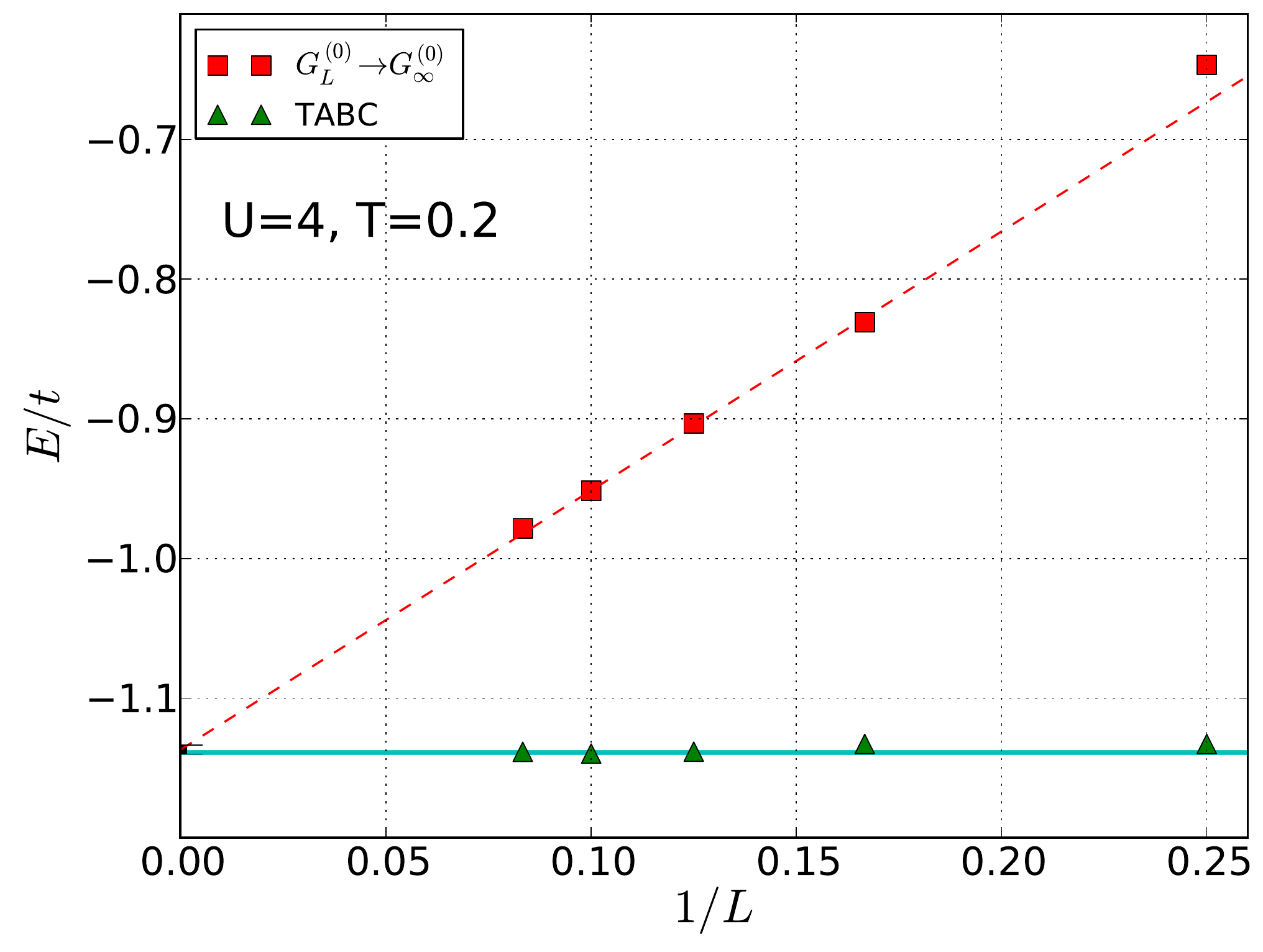}
\caption{(Color online.) Example of the dependence of energy on the inverse of the linear system size $L$ obtained with the periodic boundary conditions (PBC, circles) for $U=8$, $T=0.3875$ and with the twist-averaged boundary conditions (TABC, triangles) for $U=4$, $T=0.2$ compared to the result of a simulation based on free-particle propagators of an infinite system ($G_L^{(0)} \to G_\infty^{(0)}$, squares). The error bars are smaller than the symbols. The dashed line is a linear fit yielding $E''(\infty)=-0.5962(9)$ ($E''(\infty)=-1.1368(32)$) in perfect agreement with the claimed conservative estimate $E(\infty)=-0.5960(16)$ ($E(\infty)=-1.1390(9)$) for $U=8$, $T=0.3875$ ($U=4$, $T=0.2$) shown by the horizontal band. The data point for the smallest system size $L=4$ at $U=4$, $T=0.2$ obtained with $G_L^{(0)} \to G_\infty^{(0)}$ deviates from the linear scaling followed by larger systems and therefore is excluded from the fit.}
\label{fig:fss2}
\end{figure}

In the second approach, we replace the free-particle propagators $ G^{(0)} = G_L^{(0)}$ in the diagrammatic expansion, Eq.~\eqref{A_matr_el}, by those corresponding to the limit $L \to \infty$, $G_L^{(0)} \to G_\infty^{(0)}$, thereby completely eliminating the oscillations coming from the discreteness of the spectrum at the expense of giving up the PBC. The only source of systematic error in this case is the finiteness of the volume---still given by $L^3$---confining the distribution of the interaction vertices in Eq.~\eqref{Sp}. In this case, the finite-size corrections are substantially larger than those of simulations with PBC. However, the scaling of these corrections is linear in $1/L$ for all the local observables in question, which allows to perform a systematic TD-limit extrapolation. As an example, such an extrapolation for energy in comparison with the data for PBC (at $U=8$) and TABC (at $U=4$) is shown in Fig.~\ref{fig:fss2}. The TD-limit value obtained thereby is in perfect agreement with the result of simulations with the PBC and TABC. This constitutes an independent verification of the accuracy of the claimed results.

\subsection{Energy} 

\begin{figure*}
\includegraphics[width = 0.98\columnwidth,keepaspectratio=true]{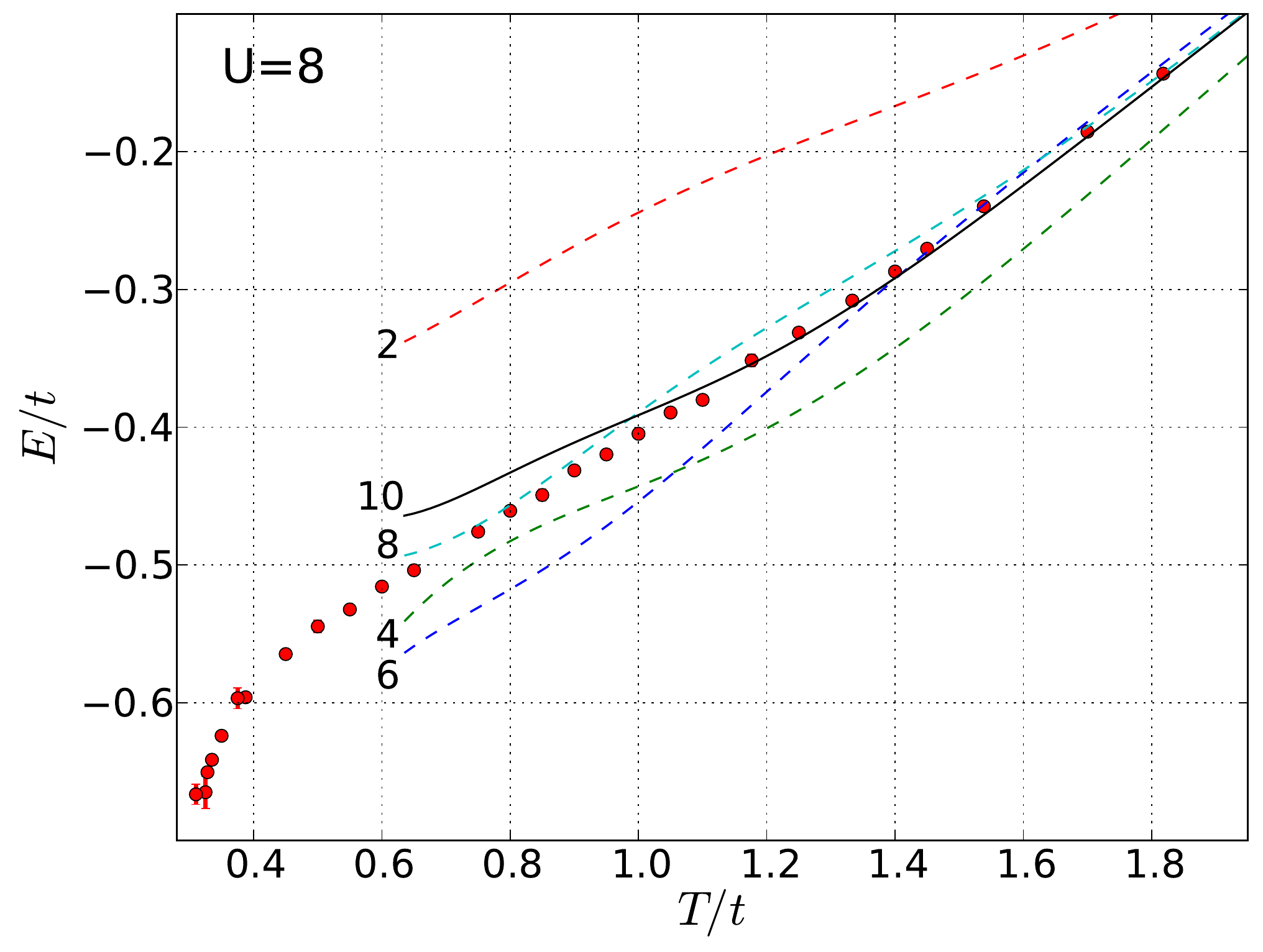}
\includegraphics[width = 0.98\columnwidth,keepaspectratio=true]{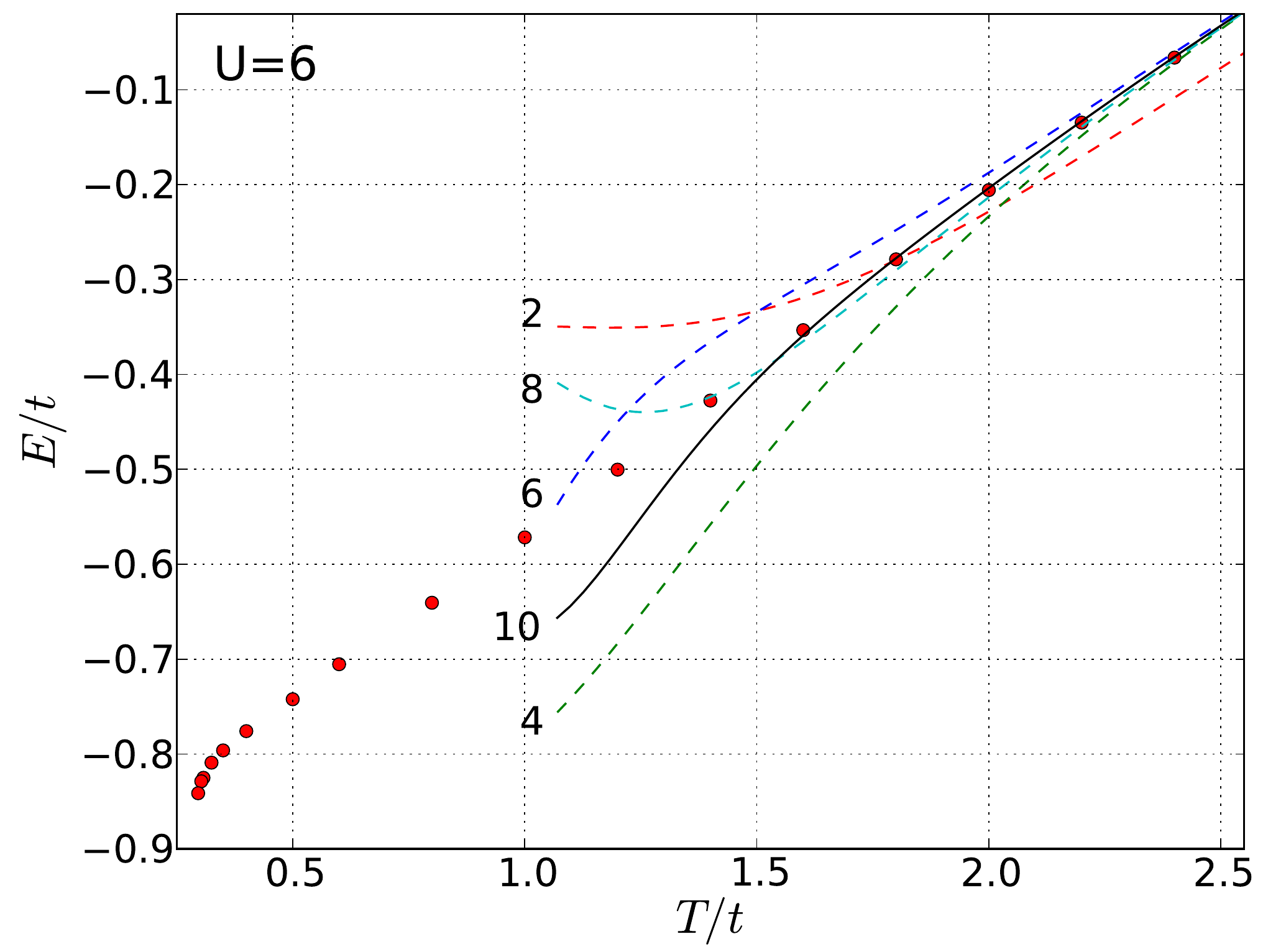}
\includegraphics[width = 0.98\columnwidth,keepaspectratio=true]{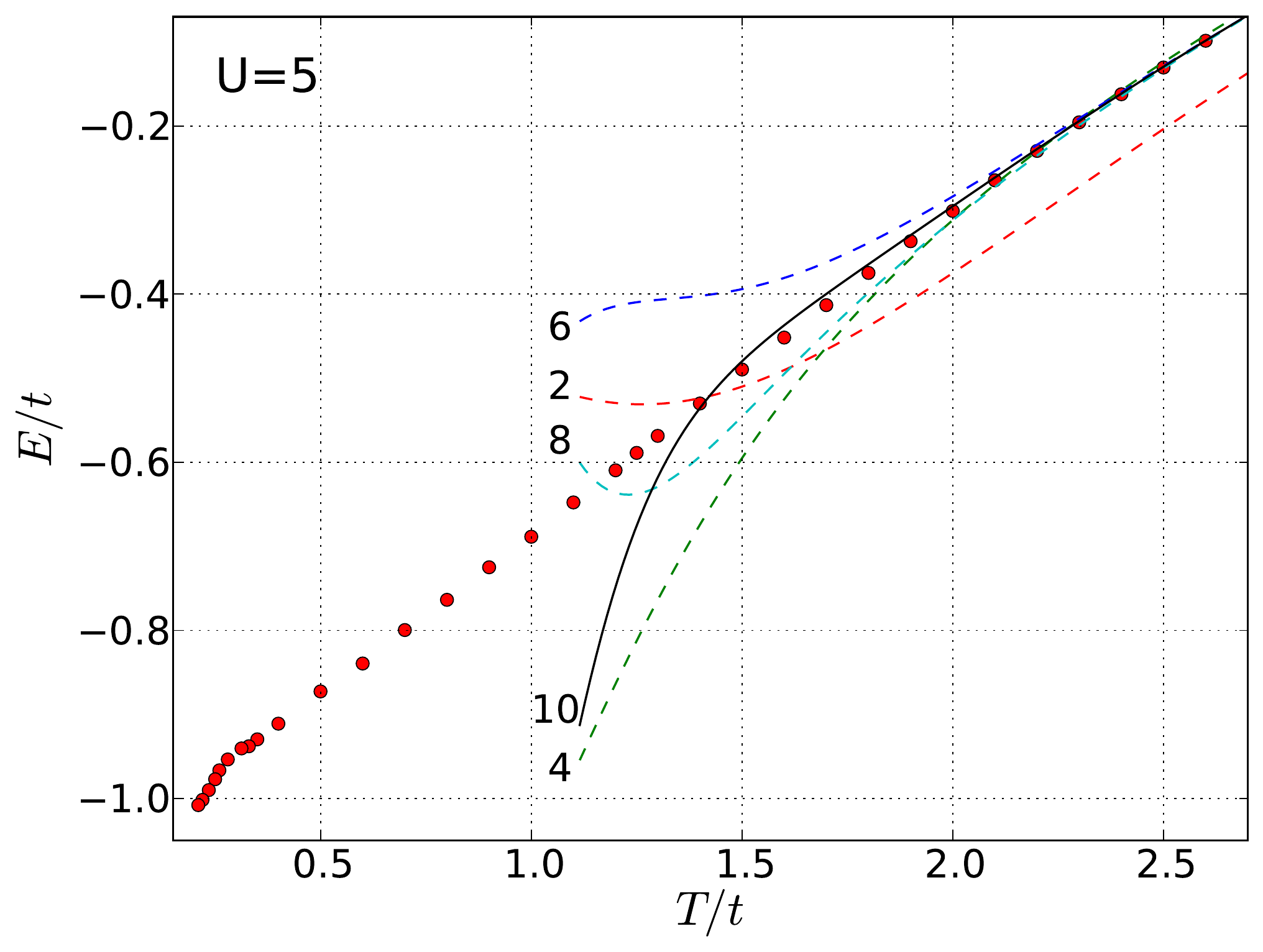}
\includegraphics[width = 0.98\columnwidth,keepaspectratio=true]{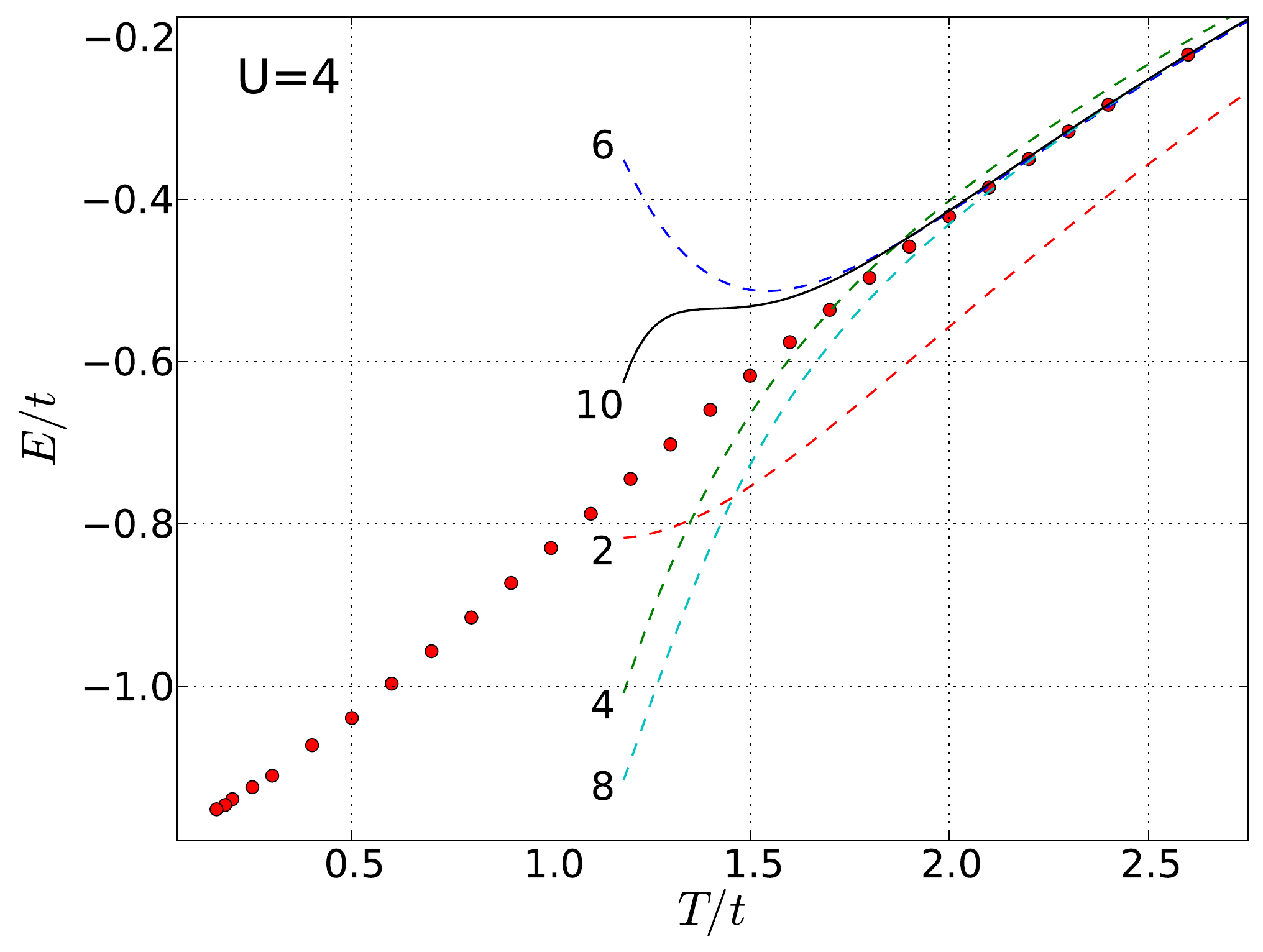}
\caption{(Color online.) Energy (extrapolated to the TD limit) versus temperature at $U=8, 6, 5, 4$. The lines represent the results of the high-temperature series expansion series of orders $2,4,6,8,10$ labeled correspondingly.}
\label{fig:En_vs_T}
\end{figure*}

Here we present simulation results for the total energy per particle extrapolated to the TD limit for a range of the interaction $U$ in the correlated regime near the N\'eel transition. The temperature dependence of the energy per particle for $U=8, 6, 5, 4$ is plotted in Fig.~\ref{fig:En_vs_T} along with the results of the HTSE \cite{Oitmaa} for orders $2,4,6,8,10$. As seen from the plot, the HTSE starts diverging well above the transition point. 

\subsection{Entropy} \label{subsec:entropy} 

The entropy per particle $S(T)$ at a given temperature $T$ is obtained from the thermodynamic relation $TdS=dE$ at fixed volume by the integral 
\begin{equation}
S(T)=S(T_*) + \frac{E(T)}{T} - \frac{E(T_*)}{T_*} - \int_{T}^{T_*} \frac{E(T')}{T'^{2}} dT', \label{entropy_integral}  
\end{equation}
where $T_*$ is some temperature at which the entropy is known. We choose $ T_* $ to be the lowest temperature at which the HTSE for the energy obviously converges to the TD-limit value $E(T_*)$ from the simulation. From Fig.~\ref{fig:En_vs_T}, we find $T_*=1.8, \; 2.4, \; 2.6, \; 2.6  $ at $U=8,\, 6, \, 5, \, 4$ respectively. Then, the accurate value of $S(T_*)$ in Eq.~\eqref{entropy_integral} is given by the HTSE, while the integral is done over the simulation data after taking the TD limit. Since the dependence $E(T)$ is slow, we represent it by a piecewise linear function and take the integral analytically. The systematic error of integration is included in the error bars for $S(T)$, but is negligible compared to the error propagated from the values of $E(T)$.

\begin{figure*}
\includegraphics[width = 0.98\columnwidth,keepaspectratio=true]{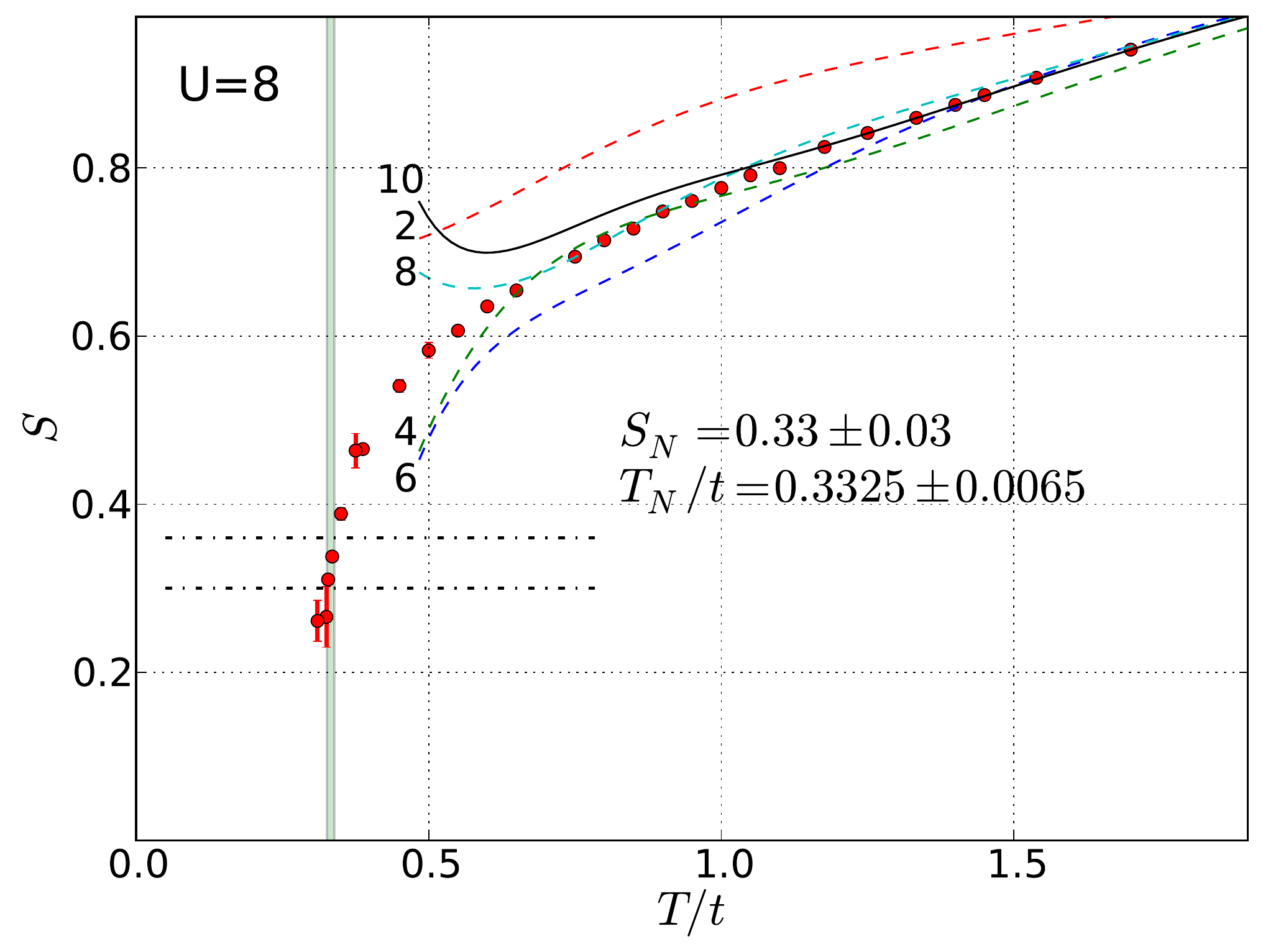}
\includegraphics[width = 0.98\columnwidth,keepaspectratio=true]{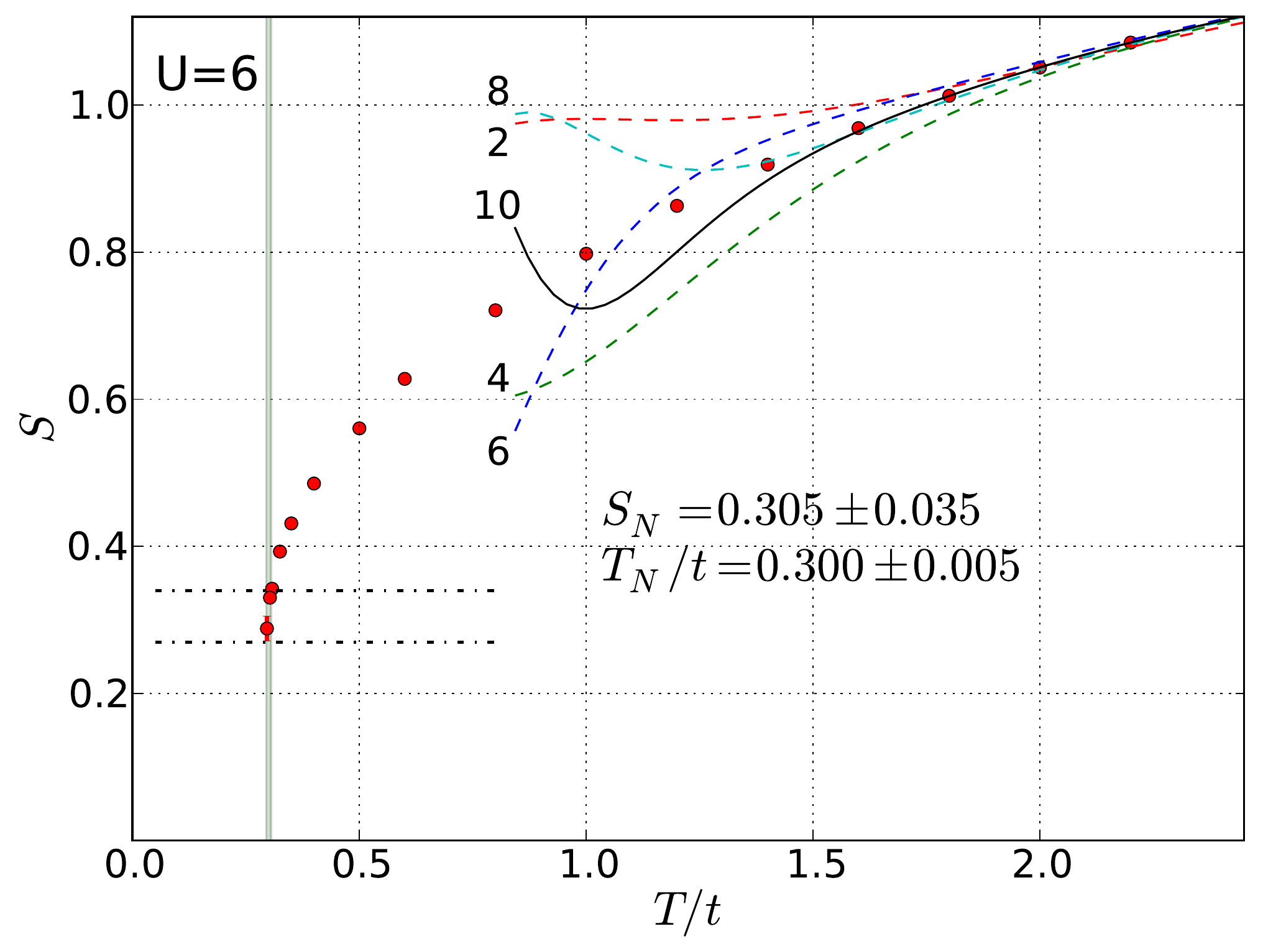}
\includegraphics[width = 0.98\columnwidth,keepaspectratio=true]{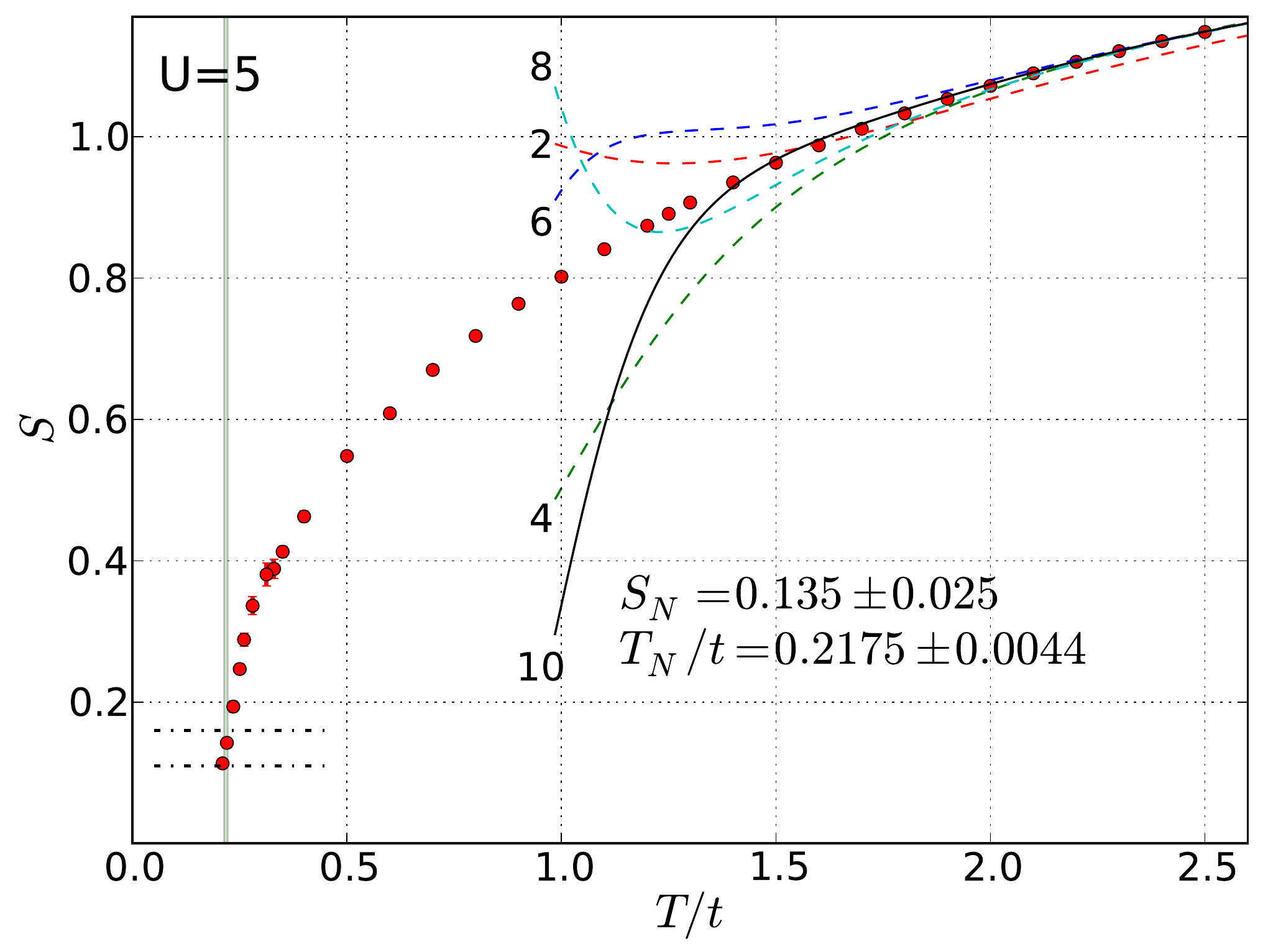}
\includegraphics[width = 0.98\columnwidth,keepaspectratio=true]{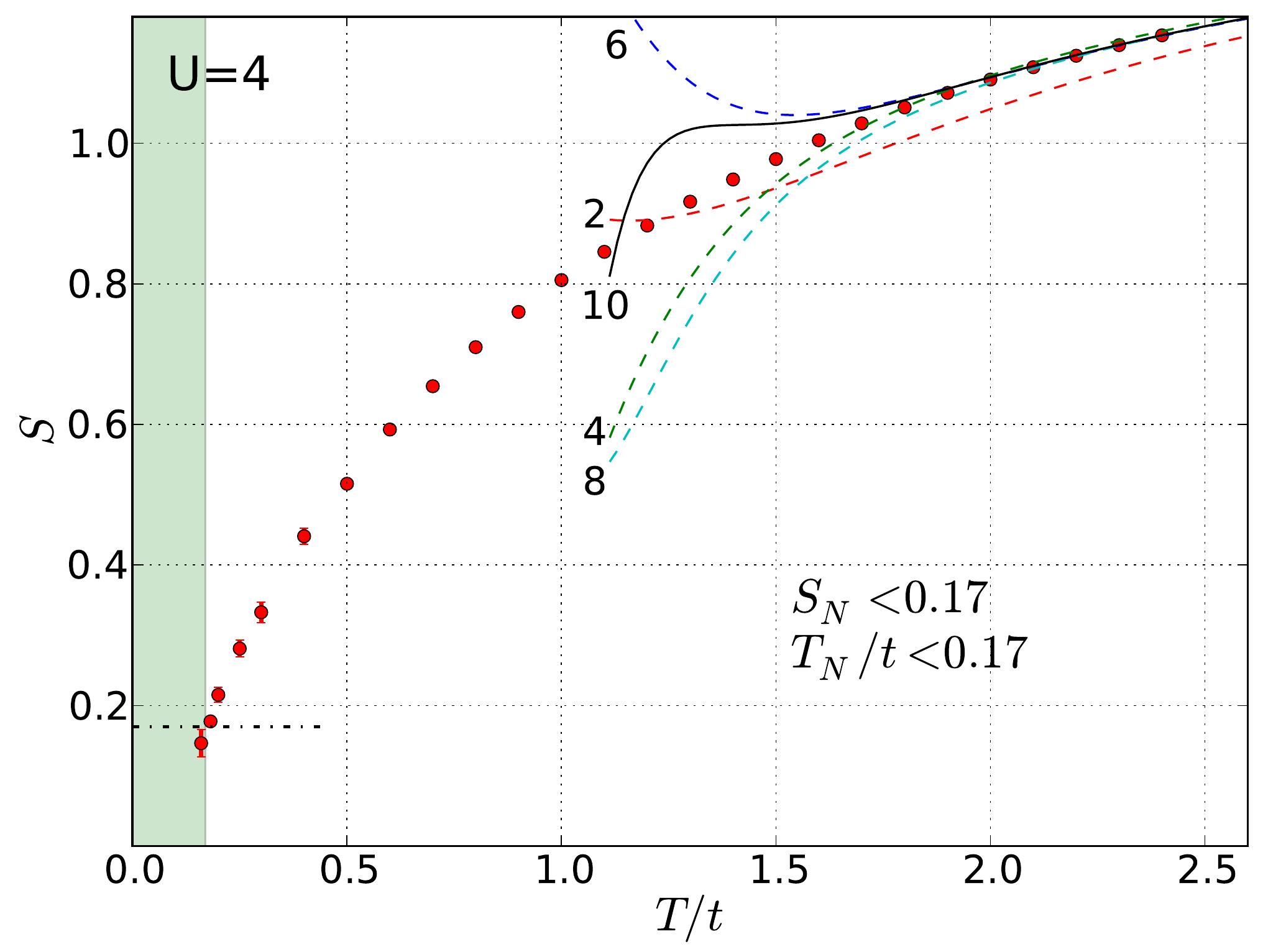}
\caption{(Color online.) Entropy (extrapolated to the TD limit) versus temperature at $U=8,6,5,4$. In each panel, the vertical line shows the position of the critical temperature $T_N$ with its width given by the error bar, while the horizontal dashed lines represent the corresponding bounds of the critical entropy $S_N$ listed in Table~\ref{table:TNs}. The rest of the lines represent the results of the high-temperature series expansion series of orders $2,4,6,8,10$ labeled correspondingly.}
\label{fig:Entr_vs_T}
\end{figure*}

\begin{figure}
\includegraphics[width = 0.98\columnwidth,keepaspectratio=true]{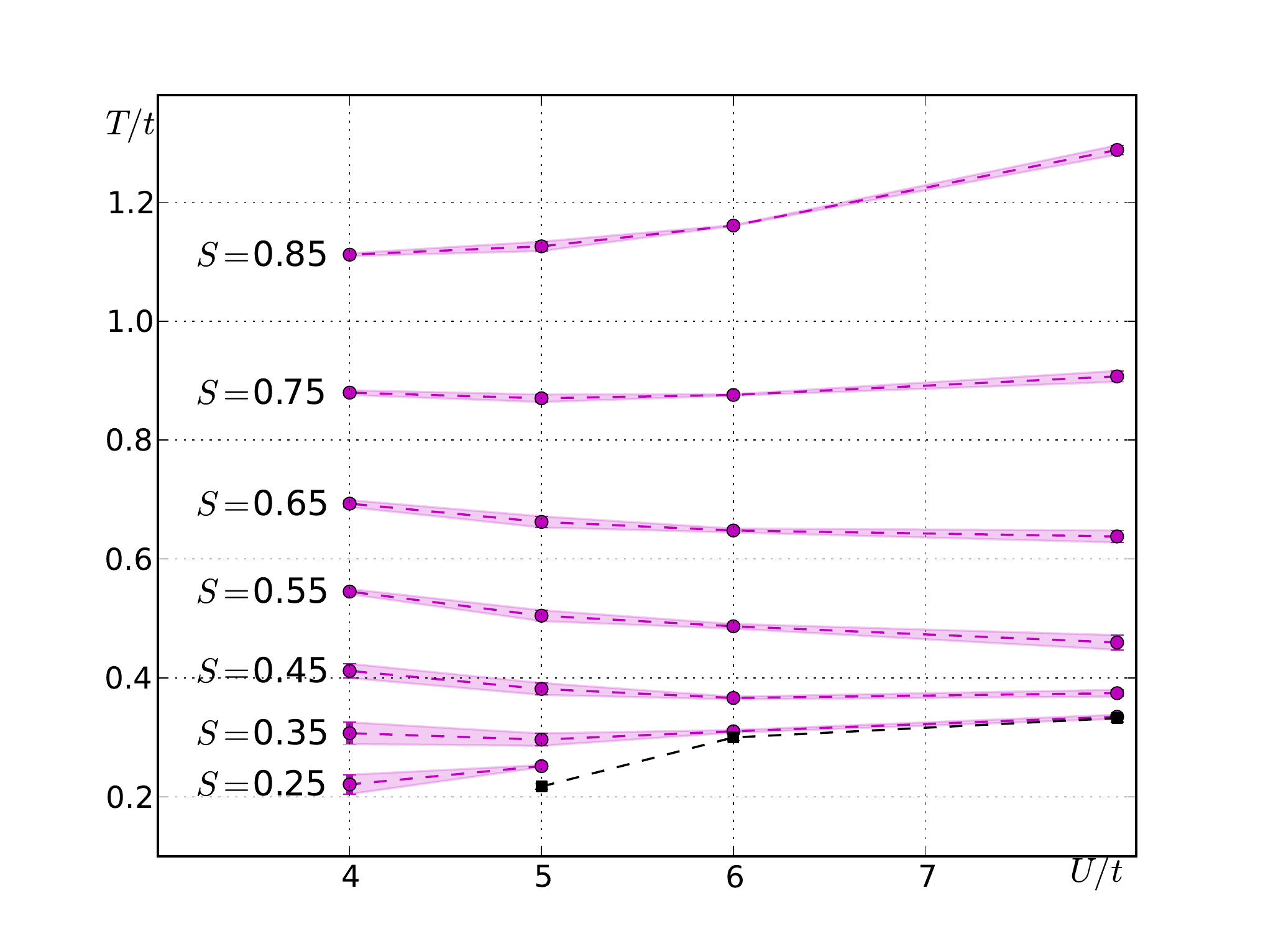}
\caption{(Color online.) Lines of constant entropy in the $T$ vs $U$ plane. }
\label{fig:T_vs_U_S=const}
\end{figure}

The resulting curves of $S(T)$ for $U=8,6,5,4$ are shown in Fig.~\ref{fig:Entr_vs_T}. From these data and our calculation of $T_N$ discussed in Sec.~\ref{sec:T_N}, we find the values of the critical entropy $S_N=S(T_N)$ in the range of $U$ and summarize the results in Table~\ref{table:TNs}. The error bars of  $S_N$ are dominated by the relatively small error of $T_N$ due to the large slope of $S_N(T)$ near the transition. In Fig.~\ref{fig:T_vs_U_S=const}, we plot lines of constant entropy in the ($T$, $U$) plane. The latter demonstrate that an adiabatic increase of the coupling $U$ can lead to either a rise (at $S \lesssim 0.35$ and $S \gtrsim 0.7$) or a fall (at $0.35  \lesssim S \lesssim 0.7$) of temperature, although the net effect of the Pomeranchuk cooling near $S_N$ is rather small.  

\subsection{Thermometry} \label{subsec:thermometry} 

As was shown in Ref.~\onlinecite{Fuchs} by means of DCA calculations, the nearest-neighbor spin-spin correlation function defined as $\langle S^{z}_\mathbf{x} S^{z}_{\mathbf{x} + \mathbf{e}_i} \rangle$, which is accessible in present-day ultracold-atom experiments, can serve as a sensitive thermometer at temperatures near $T_N$. In contrast, another routinely measured correlator, the double occupancy $\langle n_{\mathbf{x}\uparrow} n_{\mathbf{x}\downarrow} \rangle$ of a lattice site, is nearly flat in this temperature range making it a rather poor candidate for thermometry. This is hardly surprising since the latter is concerned with correlations in the charge channel, whereas the relevant physics at these temperatures is that of developing short-range spin correlations. 

\begin{figure*}[h!]
\includegraphics[width = 0.98\columnwidth,keepaspectratio=true]{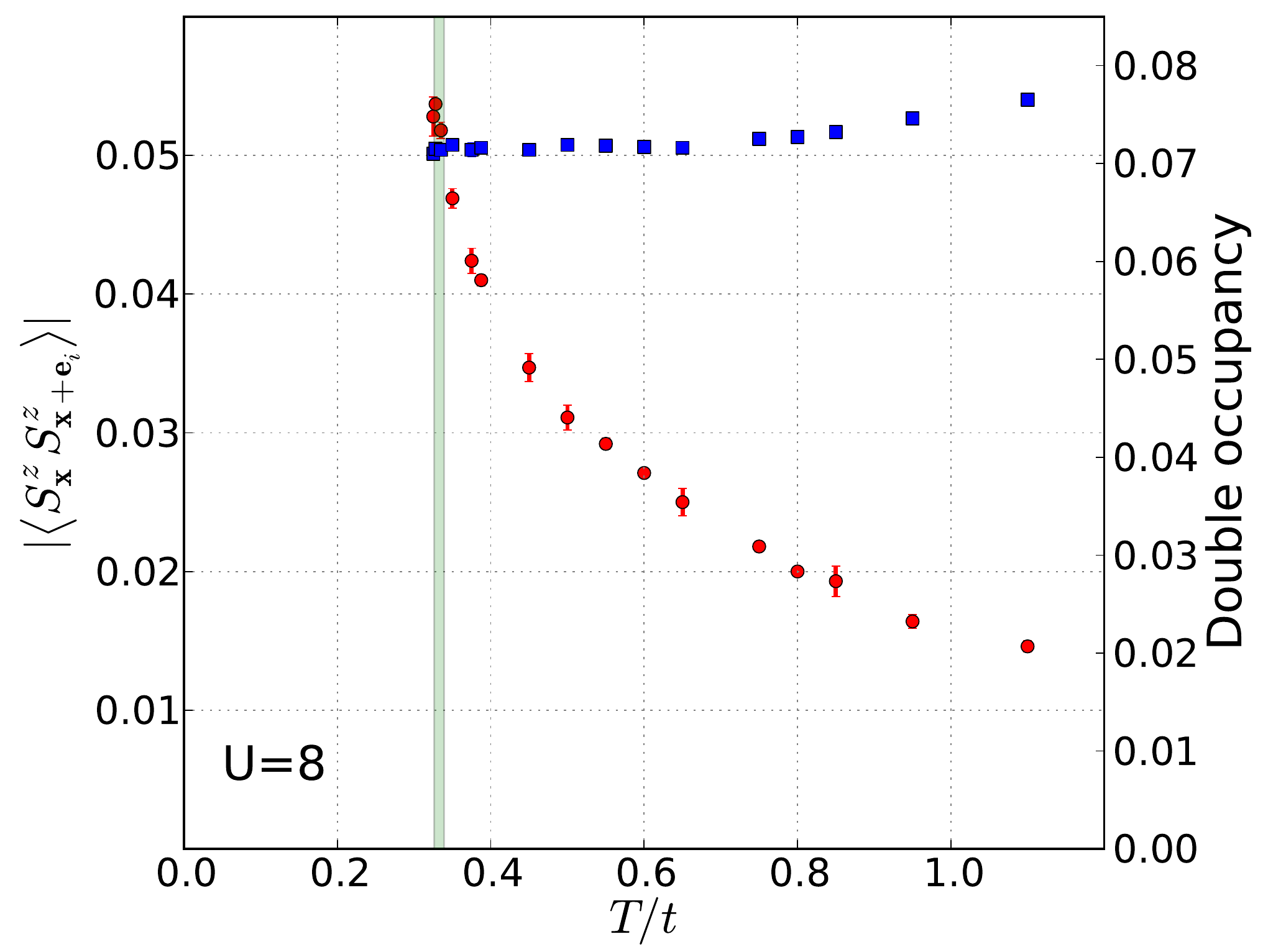}
\includegraphics[width = 0.98\columnwidth,keepaspectratio=true]{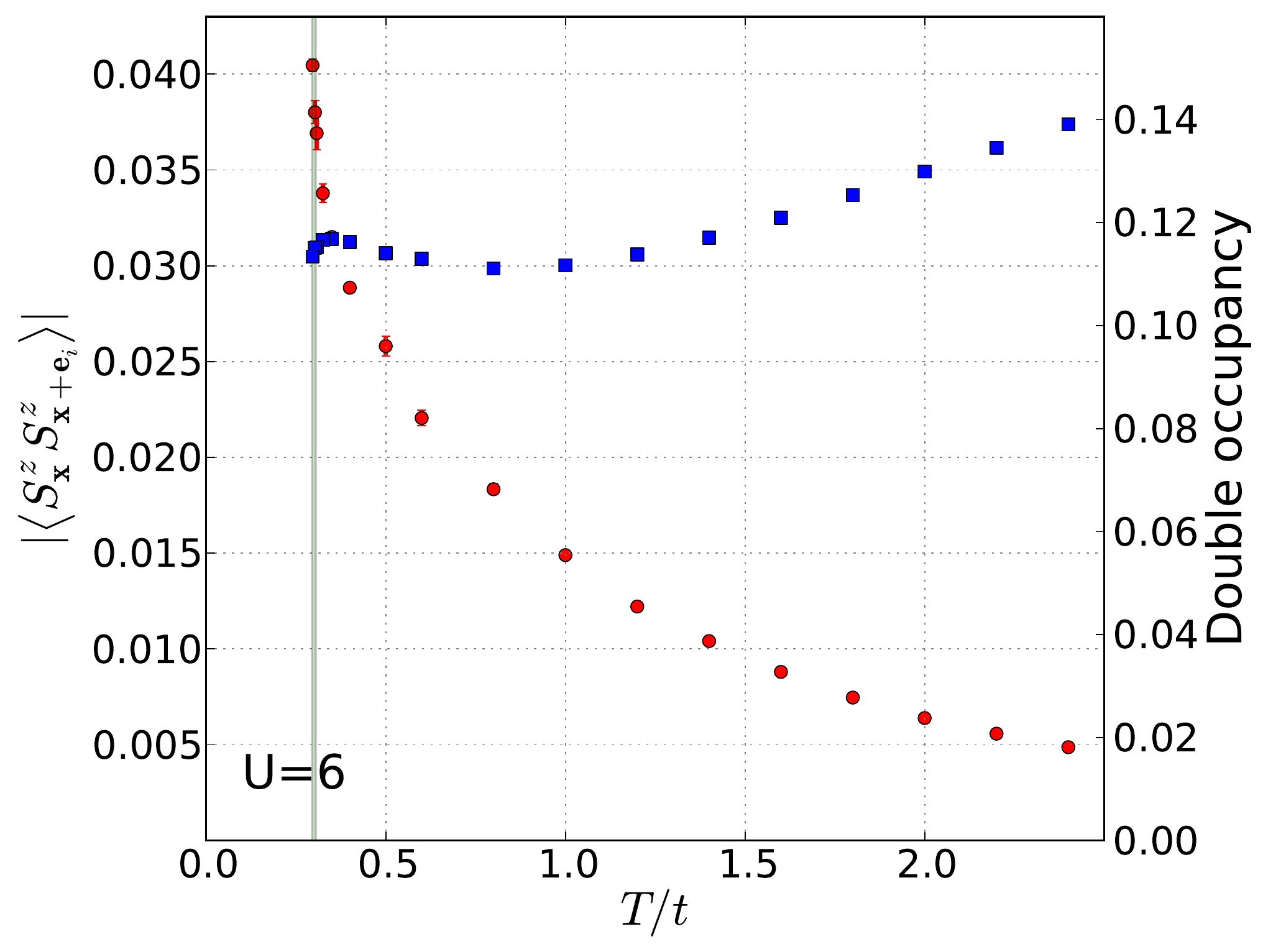}
\includegraphics[width = 0.98\columnwidth,keepaspectratio=true]{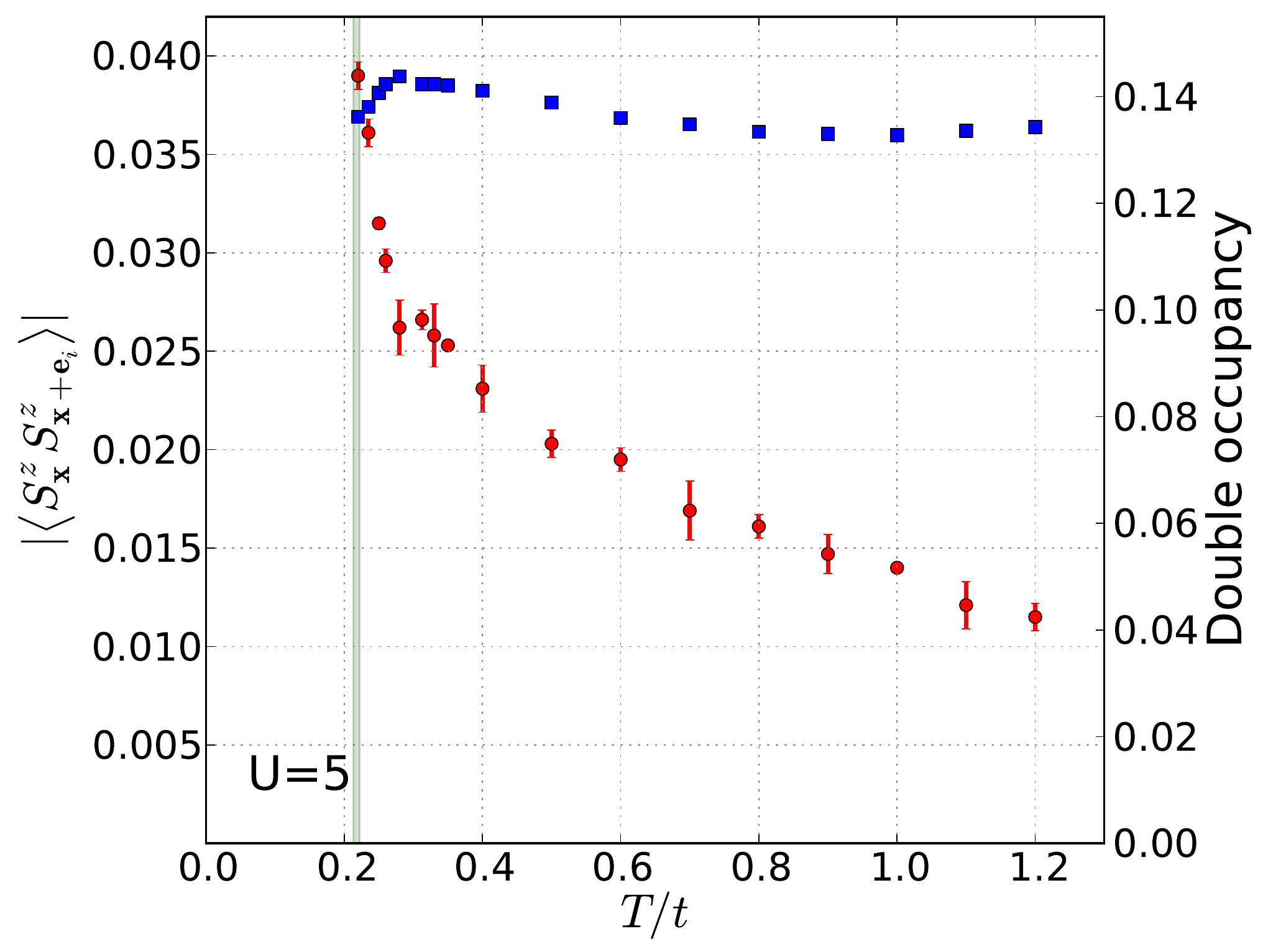}
\includegraphics[width = 0.98\columnwidth,keepaspectratio=true]{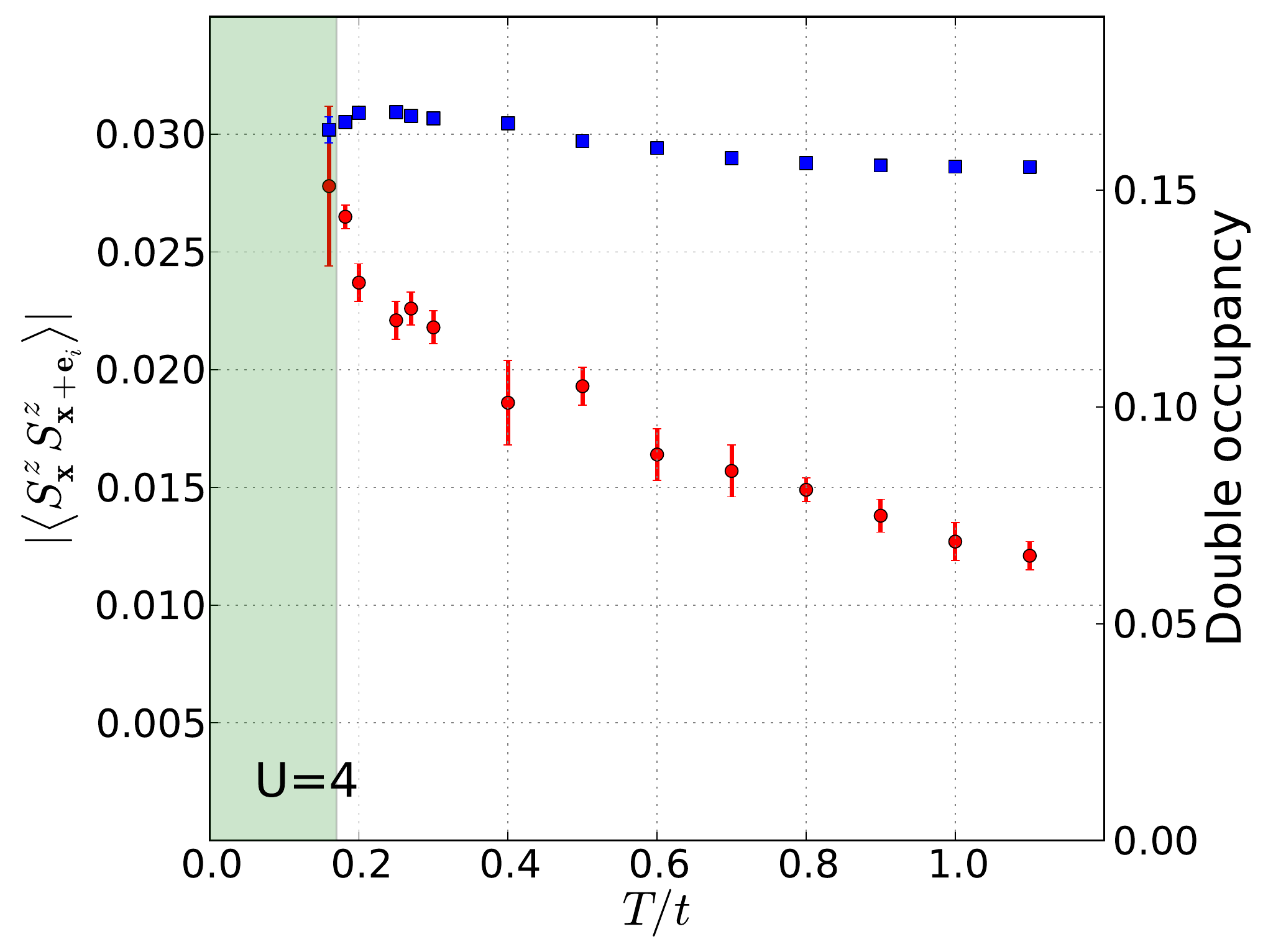}
\caption{(Color online.) Double occupancy $\langle n_{\mathbf{x}\uparrow} n_{\mathbf{x}\downarrow} \rangle$ (squares) and the nearest-neighbor spin correlation function $|\langle S_{\mathbf{x}}^{z} S_{\mathbf{x}+\mathbf{e}_i}^{z} \rangle|$ (circles) versus $T$ at $U=8,6,5,4$ extrapolated to the TD limit. In each panel, the vertical line shows the position of the critical temperature for a given value of $U$ listed in Table~\ref{table:TNs} with its width corresponding to the error bar. }
\label{fig:D_SzSz_vs_T}
\end{figure*}

In Fig.~\ref{fig:D_SzSz_vs_T}, we present our results for $\langle S^{z}_\mathbf{x} S^{z}_{\mathbf{x} + \mathbf{e}_i} \rangle$ and $\langle n_{\mathbf{x}\uparrow} n_{\mathbf{x}\downarrow} \rangle$ extrapolated to the TD limit at $U=8,6,5,4$. The obtained values agree within the errors with the TD-limit-extrapolated data from the DCA simulations, Ref.~\onlinecite{Fuchs}, but our error bars are notably smaller. Our data can be directly used for thermometry calibration and detection of the N\'eel transition. Note that in the range of temperatures $T_N<T<2T_N$ (the position of $T_N$ is depicted by a vertical line with the width corresponding to the error bar) the spin-spin correlations between nearest-neighbor cites rise by a factor of two, with a substantial increase of the slope close to $T_N$. In the same temperature range the double occupancy varies by less than $10\%$.

%
%
%

\section{Conclusions} \label{sec:conclusions} 


We presented unbiased results for the 3D Hubbard model at half filling near the N\'eel transition in the strongly correlated regime of $4 \leq U \leq 8$, where $T_N$ reaches its maximum. We focused on the properties of the model near the transition accurately determining $T_N$ and studying the energy, entropy, double occupancy  $\langle n_{\mathbf{x}\uparrow} n_{\mathbf{x}\downarrow} \rangle$ and the nearest-neighbor spin-spin correlator $\langle S^{z}_\mathbf{x} S^{z}_{\mathbf{x} + \mathbf{e}_i} \rangle$ as functions of temperature and interaction. Accurate quantitative understanding of the model in this regime is of growing importance in view of the ongoing experimental effort to emulate the Hubbard model with ultracold atoms in optical lattices, which could ultimately allow to study regions of the phase diagram inaccessible by unbiased theoretical methods. In particular, this could lead to answers of fundamental questions regarding the nature of superfluidity at finite doping and its connection to high-temperature superconductors \cite{hofstettar2002}. The realization of the N\'eel state would be a necessary step on the way to accessing the region of the phase diagram where quantum fluctuations play an important role. Our simulations provide the most accurate and controlled estimates of entropy at the critical point to date.  These entropies,  summarized in Table~\ref{table:TNs}, have to be achieved in the middle of the trapped cold-atom system to realize the AFM state. For independent \textit{in situ} thermometry in this regime, one can employ measurements of the nearest-neighbor spin correlations, which expectedly have pronounced temperature dependence near $T_N$, and which can nowadays be addressed either by the use of superlattices \cite{Bloch2010} or by lattice modulation \cite{Esslinger2011}. In agreement with Ref.~\onlinecite{Fuchs}, we did not find the double occupancy to display notable temperature dependence in the regime of interest. More generally, our results for thermodynamics at half filling quantitatively agree with the extrapolated DCA data of Ref.~\onlinecite{Fuchs} and Ref.~\onlinecite{Loh} with the combined errorbars, although the energy and entropy in DCA~\cite{Fuchs} appear to be systematically above our values as well as those of DQMC on approach to the critical point. As a result and due to the improved estimate of $T_N$, our value of $S_N$ at $U=8$ (0.33(3)) is below that claimed in Ref.~\onlinecite{Fuchs} (0.42(2)) suggesting agreement at the level of two combined standard deviations.   

The need for a more precise knowledge of $T_N$ comes from the steep temperature dependence of the entropy close to the transition. Our results for $T_N$ improve on the earlier studies of Staudt \textit{et al.} \cite{Staudt2000}, although remain in perfect agreement with the latter within the error bars everywhere but at $U=4$, where we were able to find only the upper bound for $T_N$, which is somewhat lower than the result of Ref.~\onlinecite{Staudt2000}. The results of our simulations can be used as benchmarks for tuning approximate methods as well as in developing new unbiased techniques.

The simulations were carried out on the Brutus cluster at ETH Zurich.  
E.K. acknowledges financial support of the Fellowship for Advanced Researchers by the Swiss National Science Foundation.  
E. B. gratefully acknowledges the hospitality of \textit{Laboratoire de Physique Th\'{e}orique et Mod\`{e}les Statistiques}, where a part of this work was done.  V.S.  acknowledges support from the AFOSR (FA9550~-11-1-0313) and DARPA-YFA (N66001~-11-1-4122).


\newpage
\section{Appendix} \label{appendix} 

\begin{table}[h]
\begin{tabular}{lllll}
$T/t$ & $\;\;\;\;\;\;\;  E/t \; \;\;\;\;\;\;\;\;\;$ & $\;\;\;\;\;\;\; S \; \;\;\;\;\;\;\;\;\;$ & $\;\langle n_{\mathbf{x}\uparrow} n_{\mathbf{x}\downarrow} \rangle \;\;$ & $ \;\langle S^{z}_\mathbf{x} S^{z}_{\mathbf{x} + \mathbf{e}_i} \rangle\;\;$ \\
$1.8182$  &  $-0.1434(24)$  &   $-$ & $-$ & $-$ \\
$1.7$  & $-0.1855(15)$    &  $0.9408(16)$ & $-$ &  $-$ \\
$1.5385$ & $-0.2397(26)$ & $0.9073(22)$ & $-$ & $-$ \\ 
$1.45$ &  $-0.2705(28)$  &  $0.8867(23)$ & $-$ & $-$ \\
$1.4$ & $-0.2870(31)$  & $0.8751(26)$ & $-$ & $-$ \\
$1.3333$ & $-0.3081(17)$ &  $0.8596(19)$ & $-$ & $-$ \\
$1.25$ & $-0.3314(38)$   &  $0.8416(33)$ & $-$ & $-$ \\
$1.1765$ & $-0.3515(45)$ &  $0.8250(41)$ & $-$ & $-$ \\
$1.1$ & $-0.3802(14)$  & $0.7998(19)$ & $0.0765(1)$ & $-0.0146(4)$ \\
$1.05$ & $-0.3894(35)$  & $0.7913(36)$ & $-$ & $-$ \\
$1.0$ & $-0.4048(28)$ & $0.7762(31)$ & $-$ & $-$ \\
$0.95$ & $-0.4198(22)$  &  $0.7608(27)$ & $0.0746(3)$ & $-0.0164(5)$ \\
$0.9$ & $-0.4314(30)$  & $0.7483(36)$ & $-$ & $-$ \\
$0.85$ & $-0.4492(44)$  & $0.7279(54)$ & $0.0732(3)$ & $ -0.0193(11)$ \\
$0.8$ & $-0.4607(20)$  & $0.7140(30)$ & $0.0727(3)$ & $-0.0200(3)$ \\
$0.75$ & $-0.4758(19)$  & $0.6945(29)$ & $0.0725(2)$ & $-0.0218(4)$ \\
$0.65$ & $-0.5038(31)$  & $0.6544(51)$ & $0.0716(4)$ & $-0.0250(10)$ \\
$0.6$ & $-0.5157(21)$ & $0.6354(42)$ & $0.0717(1)$ & $-0.0271(2)$ \\
$0.55$ & $-0.5323(11)$ & $0.6065(26)$ & $0.0718(3)$ & $-0.0292(4)$ \\
$0.5$ & $-0.5446(46)  $ & $0.5830(93)$ & $0.0719(4)$ & $-0.0311(9)$ \\
$0.45$ & $-0.5647(29)$  & $0.5407(76)$ & $0.0714(5)$ & $-0.0347(10)$ \\
$0.3875$ & $-0.5960(16)$ & $0.4657(57)$ & $0.0716(2)$ & $-0.0410(4)$ \\
$0.375$ & $-0.5967(76) $ & $0.464(20)$ & $0.0714(7)$ & $-0.0424(9)$ \\
$0.35$ & $-0.6240(19) $ & $0.3887(73)$ & $0.0719(4)$ & $-0.0469(7)$ \\
$0.335$ & $-0.6414(19) $  & $0.3379(59)$ & $0.0714(2)$ & $-0.0518(6)$ \\ 
$0.328$ & $-0.6504(19) $ & $0.3106(60)$ & $0.0715(2)$ & $-0.0537(3)$ \\
$0.325$ & $-0.665(11) $ & $0.266(36)$ & $0.0710(7)$ & $-0.0528(14)$ \\
$0.31$ & $-0.6664(74) $  & $0.261(24)$ & $-$ & $-$
\end{tabular}
\caption{$U=8$: Energy, entropy, double occupancy, and the nearest-neighbor spin-spin correlator as functions of temperature extrapolated to the TD limit using PBC data.}
\label{table:TDU8}
\end{table}

\begin{table}
\begin{tabular}{lllll}
$T/t$ & $\;\;\;\;\;\;\;  E/t \; \;\;\;\;\;\;\;\;\;$ & $\;\;\;\;\;\;\; S \; \;\;\;\;\;\;\;\;\;$ & $\;\langle n_{\mathbf{x}\uparrow} n_{\mathbf{x}\downarrow} \rangle \;\;$ & $ \;\langle S^{z}_\mathbf{x} S^{z}_{\mathbf{x} + \mathbf{e}_i} \rangle\;\;$ \\
$2.4$  &   $-0.0661(1)$ &   $ - $ & $0.13908(2)$ & $-0.00487(3) $ \\
$2.2$   &  $-0.1346(1)$ &   $1.0847(1)$ & $0.13452(3)$ & $-0.00558(3)$ \\
$2.0$   &  $-0.2057(2)$ &   $ 1.0508(1) $ & $0.12990(4)$ & $-0.00639(6) $ \\
$1.8$  &  $-0.2787(2)$ &   $ 1.0124(1)$ & $0.12533(4)$ & $-0.00746(5) $ \\
$1.6$   &  $-0.3533(2)$ &   $0.9685(1) $ & $0.12093(4)$ & $-0.00880(6) $ \\
$1.4$   &  $-0.4275(3)$ &   $0.9189(2) $ & $0.11705(4)$ & $-0.01041(8) $ \\
$1.2$   &  $-0.5003(4)$ &   $0.8628(3) $ & $0.11380(5)$ & $-0.01221(13) $ \\
$1.0$   &  $-0.5717(6)$ &   $0.7977(6) $ & $0.11170(7)$ & $-0.01489(19) $ \\
$0.8$   &  $-0.6406(6)$ &   $0.7208(8)$ & $0.11108(7)$ &         $-0.01833(30)$ \\
$0.6$   &  $-0.7054(15)$  &   $0.6276(26) $ & $0.11298(25)$ & $-0.02205(40)$ \\
$0.5$   &  $-0.7423(15)$  &   $0.5604(35)$ & $0.11403(29)$ &  $-0.02580(51)$ \\
$0.4$   &  $-0.7759(10)$  &   $0.4854(32)$ & $0.11620(10)$ &  $-0.02885(17)$ \\
$0.35$  &  $-0.7962(16)$  &   $0.4312(48)$ & $0.11685(25)$ &   $-0.03149(26)$ \\
$0.325$ &  $-0.8091(17)$  &   $0.3929(54) $ & $0.11662(40)$ &  $-0.03378(49)$ \\
$0.3077$ & $-0.8251(20)$  &   $0.3423(66)$ & $0.11521(50) $ & $-0.03691(87)$ \\
$0.3030$ & $-0.8288(20)$  &   $0.3303(67)$ & $0.11510(40) $ & $-0.03800(60)$ \\
$0.2963$ &  $-0.8414(50)$  &   $0.288(17) $ & $0.1134(10) $ &   $-0.04046(34)$
\end{tabular}
\caption{$U=6$: Energy, entropy, double occupancy, and the nearest-neighbor spin-spin correlator as functions of temperature extrapolated to the TD limit using PBC data.}
\label{table:TDU6}
\end{table}

\begin{table}
\begin{tabular}{lllll}
$T/t$ & $\;\;\;\;\;\;\;  E/t \; \;\;\;\;\;\;\;\;\;$ & $\;\;\;\;\;\;\; S \; \;\;\;\;\;\;\;\;\;$ & $\;\langle n_{\mathbf{x}\uparrow} n_{\mathbf{x}\downarrow} \rangle \;\;$ & $ \;\langle S^{z}_\mathbf{x} S^{z}_{\mathbf{x} + \mathbf{e}_i} \rangle\;\;$ \\
$2.6$  &  $-0.0987(6)$ &   $ -  $ & $-$ & $-$ \\
$2.5$  & $-0.1305(4)$&    $1.1483(3)$ & $-$ & $-$ \\
$2.4$  & $-0.1623(5)$&   $1.1354(3)$ & $-$ & $-$ \\
$2.3$  & $-0.1957(5) $&  $1.1212(3)$ & $-$ & $-$ \\
$2.2$  &  $-0.2297(7)$&   $1.1060(4)$ & $-$ & $-$ \\
$2.1$   & $-0.2647(10)$& $1.0898(5)$ & $-$ & $-$ \\
$2.0$   & $-0.3011(6)$&    $1.0720(3)$ & $-$ & $-$ \\
$1.9$    &$-0.3371(13)$&  $1.0535(7)$ & $-$ & $-$ \\ 
$1.8$    &$-0.3749(15)$&   $1.0331(8)$ & $-$ & $-$ \\
$1.7$    &$-0.4132(16)$&   $1.0112(9)$ & $-$ & $-$ \\
$1.6$    &$-0.4517(17)$&   $0.9879(11)$ & $-$ & $-$ \\
$1.5$    &$-0.4897(17)$&   $0.9633(12)$ & $-$ & $-$ \\
$1.4$    &$-0.5301(9)$&     $0.9354(7)$ & $-$ & $-$ \\
$1.3$    &$-0.5687(22)$&   $0.9069(16)$ & $-$ & $-$ \\
$1.25$   &$-0.5889(22)$&   $0.8910(18)$ & $-$ & $-$ \\
$1.2$    &$-0.6096(14)$&    $0.8741(12)$ & $0.1343(1)$ & $-0.0115(7)$ \\
$1.1$    &$-0.6478(32)$&    $0.8409(29)$ & $0.1336(3)$ & $-0.0121(12)$ \\
$1.0$    &$-0.6886(4)$&      $0.8020(12)$ & $0.1328(1)$ & $ -0.0140(1)$ \\
$0.9$    &$-0.7250(36)$&    $0.7637(40)$ & $0.1330(5)$ & $-0.0147(10)$ \\
$0.8$    &$-0.7637(22) $&   $0.7181(33)$ & $0.1334(4)$ &    $-0.0161(6)$ \\ 
$0.7$    &$-0.7996(41) $&   $0.6701(61)   $ & $0.1348(6)$ & $-0.0169(15)$ \\
$0.6$    &$-0.8394(15) $&   $0.6087(45)   $ & $0.1360(3)$ & $-0.0195(6)$ \\
$0.5$    &$-0.8726(11) $&   $0.5482(30)   $ & $0.1389(3)$ & $-0.0203(7)$ \\ 
$0.4$    &$-0.9110(31)$&    $0.4626(80)   $ & $0.1411(8)$ & $-0.0231(12)$ \\
$0.35$   &$-0.9296(17) $&   $0.4128(81)   $ & $0.1421(4)$ & $-0.0253(1)$ \\
$0.33$   &$-0.9379(45)$&    $0.389(14)   $ &   $0.1423(12)$ & $-0.0258(16)$ \\
$0.3125$ & $-0.9404(50)$&   $0.381(17)   $ & $0.1423(6)$ &   $-0.0266(5)$ \\  
$0.28$  & $-0.9535(30)  $&   $0.336(13)   $ &   $0.1438(8)$ &  $-0.0262(14)$ \\
$0.26$  & $-0.9665(21) $&    $0.2884(94)   $ & $0.1423(4)$ &  $-0.0296(6)$ \\
$0.25$  & $-0.9771(19) $&    $0.2468(78)   $ & $0.1407(10)$ & $-0.0315(3)$ \\
$0.235$ & $-0.9900(18)$&    $0.1936(77)   $ & $0.1381(10)$ & $-0.0361(7) $ \\
$0.22$  & $-1.0016(16)$&     $0.1424(76)   $ & $0.1362(11)$ & $-0.0390(7)$ \\
$0.21$  & $-1.0079(10)$&     $0.1133(53)   $ & $-$ & $-$ 
\end{tabular}
\caption{$U=5$: Energy, entropy, double occupancy, and the nearest-neighbor spin-spin correlator as functions of temperature extrapolated to the TD limit using TABC data.}
\label{table:TDU5}
\end{table}

\begin{table}
\begin{tabular}{lllll}
$T/t$ & $\;\;\;\;\;\;\;  E/t \; \;\;\;\;\;\;\;\;\;$ & $\;\;\;\;\;\;\; S \; \;\;\;\;\;\;\;\;\;$ & $\;\langle n_{\mathbf{x}\uparrow} n_{\mathbf{x}\downarrow} \rangle \;\;$ & $ \;\langle S^{z}_\mathbf{x} S^{z}_{\mathbf{x} + \mathbf{e}_i} \rangle\;\;$ \\
$2.6$ & $-0.2218(2)$ &  $-$ & $-$ & $ $ \\
$2.4$  &$-0.2837(3)$ &  $1.1534(1)$ & $-$ & $-$ \\
$2.3$  &$-0.3164(5)$ &  $1.1395(2)$ & $-$ & $-$ \\
$2.2$  &$-0.3504(3)$ &  $1.1244(1)$ & $-$ & $-$ \\
$2.1$  &$-0.3854(6)$ &  $1.1081(2)$ & $-$ & $-$ \\
$2.0$  &$-0.4213(3)$ &  $1.0906(2)$ & $-$ & $-$ \\
$1.9$  &$-0.4583(4)$ &  $1.0716(2)$ & $-$ & $-$ \\
$1.8$  &$-0.4967(4)$ &  $1.0508(2)$ & $-$ & $-$ \\
$1.7$  &$-0.5363(6)$ &  $1.0282(4)$ & $-$ & $-$ \\
$1.6$  &$-0.5759(4)$ &  $1.0042(3)$ & $-$ & $-$ \\
$1.5$  &$-0.6174(7)$ &  $0.9774(5)$ & $-$ & $-$ \\  
$1.4$  &$-0.6594(5)$ &  $0.9484(4)$ & $-$ & $-$ \\
$1.3$  &$-0.7021(8)$ &  $0.9168(6)$ & $-$ & $-$ \\ 
$1.2$  &$-0.7445(4)$ &     $0.8829(4)$ &   $-$ & $-$ \\
$1.1$  &$-0.7875(12)$ &  $0.8455(11)$ & $0.1553(2)$ & $-0.0121(6)$ \\
$1.0$   &$-0.8296(17)$ &  $0.8053(17)$ & $0.1554(2)$ & $-0.0127(8)$ \\
$0.9$   &$-0.8727(20)$ &  $0.7599(24)$ & $0.1557(5)$ & $-0.0138(7)$ \\  
$0.8$   &$-0.9152(15)$ &  $0.7098(21)$ & $0.1562(3)$ & $-0.0149(5)$ \\
$0.7$   &$-0.9568(29)$ &  $0.6543(42)$ & $0.1574(6)$ & $-0.0157(11)$ \\
$0.6$   &$-0.9967(23)$ &  $0.5927(46)$ & $0.1597(7)$ & $-0.0164(11)$ \\
$0.5$   &$-1.0390(20)$ &  $0.5155(49)$ & $0.1613(7)$ & $-0.0193(8)$ \\
$0.4$     &  $-1.0725(44)$ &  $0.441(12)$ & $0.1654(13)$ & $-0.0186(18)$ \\
$0.3$      & $-1.1101(19)$ &  $0.333(15)$ & $0.1665(5)$ & $-0.0218(7)$ \\
$0.27$     &$-$ &  $-$ & $0.1671(9)$ & $-0.0226(7)$ \\
$0.25$     &$-1.1243(25)$ &  $0.281(12)$ & $0.1680(7)$ & $-0.0221(8)$ \\
$0.2$       &$-1.1390(9)$ &    $0.215(11)$ &   $0.1678(4)$ & $-0.0237(8)$ \\   
$0.1818$ &$-1.1462(8)$ &    $0.1775(56)$ & $0.1657(5)$ & $-0.0265(5)$ \\    
$0.16$     &$-1.1515(31)$ &  $0.147(20)$ &   $0.1639(30)$ & $-0.0278(34)$ 
\end{tabular}
\caption{$U=4$: Energy, entropy, double occupancy, and the nearest-neighbor spin-spin correlator as functions of temperature extrapolated to the TD limit using TABC data.}
\label{table:TDU4}
\end{table}

\end{document}